\documentclass[a4paper,fleqn,usenatbib]{mnras}

\usepackage{amsmath}	

\usepackage[T1]{fontenc}
\usepackage{ae,aecompl}

\usepackage{pdflscape}	
\usepackage{longtable} 
\usepackage{footnote}
\usepackage[flushleft]{threeparttable}

\usepackage{graphicx}	
\usepackage{amssymb}	
\usepackage{soul,color,float}
\usepackage{CJKutf8}

\newcommand{\lzifu} {{\scshape lzifu}}

\newcommand{\magphys} {{\scshape magphys}}

\newcommand{\kms}{km~s$^{-1}$}

\newcommand{\OI}{[\ion{O}{i}]} 
\newcommand{\OIII}{[\ion{O}{iii}]} 
\newcommand{\NII}{[\ion{N}{ii}]}
\newcommand{\SII}{[\ion{S}{ii}]}  
\newcommand{\HII}{\ion{H}{ii}} 
\newcommand{\HI}{\ion{H}{i}} 
\newcommand{\Dn}{$\rm D_n(4000)$}
\newcommand{\Hd}{${\rm H}\delta_{\rm A}$}

\def\ifa{1}
\def\anu{2}
\def\sifa{3}
\def\cfar{4}
\def\kingab{5}
\def\casstro{6}
\def\ifaue{7}
\def\aao{8}
\def\swin{9}
\def\uwa{10}
\def\env{11}
\def\macu{12}
\def\umel{13}
\def\cardiff{14}

\title[SAMI: extraplanar gas and galactic winds]{The SAMI Galaxy Survey: extraplanar gas, galactic winds, and their association with star formation history}

\author[I.-T. Ho et al.]
{I-Ting Ho (何宜庭)$^{\ifa,\anu}$\thanks{E-mail: \href{itho@ifa.hawaii.edu}{itho@ifa.hawaii.edu}}, Anne M. Medling$^\anu$, Joss Bland-Hawthorn$^\sifa$, Brent Groves$^{\anu}$, 
\newauthor Lisa J. Kewley$^{\ifa,\anu}$, Chiaki Kobayashi$^{\cfar}$, Michael A. Dopita$^{\anu,\kingab}$, Sarah K. Leslie$^\anu$, 
\newauthor Rob Sharp$^{\anu,\casstro}$, James T. Allen$^{\sifa,\casstro}$, Nathan Bourne$^\ifaue$, Julia J. Bryant$^{\sifa,\casstro,\aao}$, 
\newauthor Luca Cortese$^{\swin,\uwa}$, Scott M. Croom$^{\sifa,\casstro}$, Loretta Dunne$^\ifaue$, L. M. R. Fogarty$^{\sifa,\casstro}$, 
\newauthor Michael Goodwin$^\aao$, Andy W. Green$^\aao$, Iraklis S. Konstantopoulos$^{\aao,\env}$, 
\newauthor Jon S. Lawrence$^\aao$, Nuria P. F. Lorente$^\aao$, Matt S. Owers$^{\aao,\macu}$, Samuel Richards$^{\sifa,\casstro,\aao}$, 
\newauthor Sarah M. Sweet$^{\anu}$, Edoardo Tescari$^{\casstro,\umel}$ and Elisabetta Valiante$^{\cardiff}$\\ 
$^\ifa$Institute for Astronomy, University of Hawaii, 2680 Woodlawn Drive, Honolulu, HI 96822, USA \\
$^\anu$Research School of Astronomy and Astrophysics, Australian National University, Canberra, ACT 2611, Australia\\
$^\sifa$Sydney Institute for Astronomy, School of Physics, University of Sydney, NSW 2006, Australia\\
$^\cfar$Centre for Astrophysics Research, Science and Technology Research Institute, University of Hertfordshire, Hertfordshire AL10 9AB, UK\\
$^\kingab$Astronomy Department, King Abdulaziz University, P.O. Box 80203, Jeddah, Saudi Arabia \\ 
$^\casstro$ARC Centre of Excellence for All-sky Astrophysics (CAASTRO)\\
$^\ifaue$Institute for Astronomy, University of Edinburgh, Royal Observatory, Blackford Hill, Edinburgh, EH9 3HJ, UK\\
$^{\aao}$Australian Astronomical Observatory, PO Box 915, North Ryde, NSW 1670, Australia\\
$^{\swin}$Centre for Astrophysics \& Supercomputing, Swinburne University of Technology, Victoria, Australia\\
$^{\uwa}$International Centre for Radio Astronomy Research, University of Western Australia, 35 Stirling Highway, Crawley, WA 6009, Australia\\
$^{\env}$Envizi Group Suite 213, National Innovation Centre, Australian Technology Park, 4 Cornwallis Street, Eveleigh, NSW 2015, Australia\\
$^{\macu}$Department of Physics and Astronomy, Macquarie University, NSW 2109, Australia\\
$^{\umel}$School of Physics, University of Melbourne, Parkville, VIC 3010, Australia\\
$^{\cardiff}$School of Physics and Astronomy, Cardiff University, Queen's Buildings, Cardiff, CF24 3AA, UK }

\date{Accepted 2016 January 04.  Received 2016 January 03; in original form 2015 September 22}

\pubyear{2015}

\begin{document}
\begin{CJK*}{UTF8}{bkai}

\label{firstpage}
\pagerange{\pageref{firstpage}--\pageref{lastpage}}
\maketitle

\begin{abstract}
We investigate a sample of 40 local, main-sequence, edge-on disc galaxies using integral field spectroscopy with the Sydney-AAO Multi-object Integral field spectrograph (SAMI) Galaxy Survey to understand the link between properties of the extraplanar gas and their host galaxies. The kinematics properties of the extraplanar gas, including velocity asymmetries and increased dispersion, are used to differentiate galaxies hosting large-scale galactic winds from those dominated by the extended diffuse ionized gas. We find rather that a spectrum of diffuse gas-dominated to wind dominated galaxies exist. The wind-dominated galaxies span a wide range of star formation rates ($-1 \lesssim \log({\rm SFR/M_{\sun} yr^{-1}}) \lesssim 0.5$) across the whole stellar mass range of the sample ($8.5 \lesssim \log({\rm M_{*}/M_{\sun}}) \lesssim 11$). The wind galaxies also span a wide range in SFR surface densities ($10^{-3}\mbox{--}10^{-1.5}~\rm M_{\sun} ~yr^{-1}~kpc^{-2}$) that is much lower than the canonical threshold of $\rm0.1~M_{\sun} ~yr^{-1}~kpc^{-2}$. The wind galaxies on average have higher SFR surface densities and higher \Hd\ values than those without strong wind signatures. The enhanced \Hd\ indicates that bursts of star formation in the recent past are necessary for driving large-scale galactic winds. We demonstrate with Sloan Digital Sky Survey data that galaxies with high SFR surface density have experienced bursts of star formation in the recent past. Our results imply that the galactic winds revealed in our study are indeed driven by bursts of star formation, and thus probing star formation in the time domain is crucial for finding and understanding galactic winds. 
\end{abstract}



\section{Introduction}

In the standard picture of galaxy formation and evolution, the feedback related to processes that drive energy and momentum into the interstellar gas serve to regulate the assembly of baryonic matter in dark matter haloes. Outflows from galaxies prevent further gas accretion and eject gas, metals, and energy out to many kpc into their haloes. The overy of large amounts of gas, metals, and, dust in the circumgalactic medium (at a few hundred kpc radius) has established the important role that this halo matter must play in galaxy evolution \citep{Menard:2010fy,Tumlinson:2011lr,Werk:2013fk,Werk:2014yg,Peek:2015fp}. The halo gas may eventually cool down and fall back to the disc to feed subsequent star formation (i.e. the ``galactic fountain'' picture; \citealt{Shapiro:1976nx,Bregman:1980eu,de-Avillez:2000qf}), or it may be lost to the system through interactions with other galaxies. The interplay between the discs and their haloes can strongly influence the different pathways galaxies evolve upon over cosmic time. 

Star formation and active galactic nuclei (AGN) are the two major energy sources capable of ejecting baryonic matter from the disc into the halo. Different energy input rates and their time-scales dictate the form of interactions between galaxies and their haloes. 

At the very energetic end, powerful AGNs can deposit energy and momentum from small to large scales, ejecting multiphase gas at velocities of more than a thousand kilometres per second \citep[e.g.][]{Tremonti:2007fj,Rupke:2013qy,Veilleux:2013uq}. At intermediate energy, galactic winds driven by star formation can typically reach a speed of a few hundred kilometres per second and are known to be very common at high redshifts ($z>1$) where the cosmic star formation rates (SFRs) are high \citep[e.g.][]{Weiner:2009lr,Steidel:2010fk}. In the local Universe, starburst-driven winds are ubiquitous in galaxies with high enough SFR surface densities ($\Sigma > 0.1\rm M_{\sun}~yr^{-1}~kpc^{-2}$; \citealt{Heckman:2002fk}). Although present-day normal galaxies do not have such high SFR surface densities, the prevalence of galactic winds is still implied statistically by stacking analysis with the \ion{Na}{d} absorption lines \citep{Chen:2010qy}.  At intermediate redshifts ($0.3<z<1$), the ubiquity of galactic-scale outflows has also been inferred by absorption line studies in individual systems using 10-m class telescopes (e.g. \citealt{Rubin:2014ly,Schroetter:2015db}; see also \citealt{Sato:2009ys,Martin:2012rt,Kornei:2012zr}). Finally at low energy, the extended diffuse ionized gas (eDIG), as part of the warm ionized medium seen both in our Milky Way (also known as the Reynolds Layer; \citealt{Reynolds:1973dq}) and external galaxies represents the interface between the hot haloes and the cold discs (see \citealt{Haffner:2009fr} for an extensive review).

The interactions between disc activities and haloes (the so-called ``disc-halo interactions'') can be effectively probed by study of the extraplanar emissions at the interface where the interactions occur. Edge-on galaxies provide the best viewing angle for the investigation of the structure, excitation, and dynamics of the extraplanar gas. 

In nearby edge-on galaxies that host well-known starburst driven winds, such as M82 \citep{Axon:1978fv,Bland:1988yq,Leroy:2015yu}, NGC253 \citep{Westmoquette:2011uq}, NGC3079 \citep{Veilleux:1994hl,Cecil:2001uq,Cecil:2002vn} and NGC1482 \citep{Veilleux:2002fk}, open-ended bipolar structures can extend several kpc from the central energy injection zones undergoing concentrated starbursts. Strong supernova feedback drives multiphase gas consisting of hot $10^{7-8}$~K X-ray emitting gas, warm $10^4$~K ionized gas, and cold neutral atomic and molecular gas, to a speed of a few hundred kilometres per second (see \citealt{Veilleux:2005qy} for a review). The optical line-emitting gas usually presents well-defined conical structures, which are the limb-brightened parts of the expanding X-ray bubbles that also entrain the ambient cold gas. As the bubbles expand supersonically, shock waves excite optical emission lines and produce the characteristically large [\ion{O}{i}], [\ion{O}{ii}], [\ion{N}{ii}] and [\ion{S}{ii}] to Balmer line ratios. This extraplanar gas in nearby prototype wind systems represents a very violent form of interactions between discs and haloes.

In galaxies lacking spectacular large-scale winds, narrowband H$\alpha$ imaging in nearby late-type edge-on galaxies reveals that eDIG is very common, with more than half of the galaxies showing extraplanar diffuse emissions and sometimes filamentary structures \citep[e.g.][]{Rossa:2000qf,Rossa:2003uq,Rossa:2003fj,Miller:2003yq,Rossa:2004fk}. The average distances of the extended emission above the galactic midplane can range from 1--2 to 4 kpc or more. Kinematic studies suggest that the eDIG co-rotates with the host galaxy and usually presents a slight lag in the azimuthal velocity, which increases with increasing off-plane distance \citep{Heald:2006gf,Heald:2006ul,Heald:2007ve}. O-stars in \HII\ regions are likely to contribute significantly to the excitation of eDIG \citep{Miller:1993rb,Dove:1994xy}. Escaping Lyman continuum photons from O-star \HII\ regions travelling in a low density, fractal medium or superbubble chimney could reach kpc scales and excite the extraplanar gas. This picture naturally explains the observed correlations between the far infrared luminosity per unit area and extraplanar ionized mass \citep[and also the presence of extraplanar emission][]{Miller:2003yq,Rossa:2003uq}. This picture also explains the strong spatial correlation between strong diffuse emission and \HII\ regions in the disc \citep[][]{Zurita:2000sh,Zurita:2002ri}. However, \citet{Miller:2003vn} modelled the forbidden line ({[\ion{O}{iii}]~$\lambda\lambda$4959,5007}, {[\ion{O}{i}]~$\lambda$6300}, and {[\ion{S}{ii}]~$\lambda\lambda$6716,6731}) to Balmer line ratios of the extended filaments which typically increase with increasing height off the disc plane \citep[e.g.][]{Otte:2001lr,Otte:2002fp}. They found that in most cases photoionization is unlikely to be the only excitation mechanism, with hybrid models combining photoionization and turbulent mixing layers or shocks better explaining the optical line ratios (see also \citealt{Reynolds:1999bk,Collins:2001bj,Seon:2009le,Barnes:2014pi}).

Understanding the underlying physical processes and galaxy properties that drive the different types of disc-halo interactions is critical for building a more comprehensive picture of the cycle of baryonic matter in galaxy evolution. Extraplanar gas can shed light on the different strengths and forms of the interactions. Observing such interactions in large numbers of galaxies has been made possible recently by virtue of the growth of integral field spectroscopy. The on-going integral field spectroscopy (IFS) surveys, such as the the Calar Alto Legacy Integral Field Area Survey \citep[CALIFA;][]{Sanchez:2012fj}, the Sydney-AAO Multi-object Integral field spectrograph (SAMI) Survey \citep{Croom:2012qy,Bryant:2015bh}, and the Mapping Nearby Galaxies at Apache Point Observatory (MaNGA) Survey \citep{Bundy:2015kx}, are delivering three dimensional datacubes of hundreds to tens of thousands of galaxies and starting to revolutionize the way galaxies are studied. In this work, we will explore extraplanar emissions in edge-on discs using data from the SAMI Galaxy Survey and investigate the connection between properties of the extraplanar gas and physical properties of the galaxies. 

The paper is structured as follows.  Section~2 describes our sample selection. We uss the analysis of our IFS datacubes in Section~3, and introduce in Section~4 a novel approach of empirically identifying wind-dominated galaxies. In Section~5, we detail the derivation of host galaxy properties including star formation rate and star formation history.  In Section~6, we investigate common physical properties of wind-dominated galaxies. Finally, we uss the implications of our results in Section~7 and provide a summary and conclusion in Section~8. 

Through out this paper, we assume the concordance $\Lambda$ cold dark matter cosmology with $ H_0 = 70~\rm km~s^{-1}~Mpc^{-1}$, $\rm \Omega_M = 0.3$ and $\rm \Omega_{\Lambda} = 0.7$. We infer cosmological distances from  redshifts that have been corrected for local and large-scale flows using the flow model by \citet{Tonry:2000lr}. 

\section{Sample Selection}

We select edge-on star-forming disc galaxies from the SAMI Galaxy Survey. The on-going SAMI Galaxy Survey observes low redshift galaxies ($z\approx0.05$) using deployable imaging fiber bundles (``hexabundles'') to obtain spatially-resolved optical spectra over a circular field of view (FOV) of 15 arcsec in diameter \citep{Croom:2012qy}. The survey is conducted on the 3.9-m Anglo-Australian Telescope using the flexible AAOmega dual-beam spectrograph \citep{Smith:2004eu,Saunders:2004oq,Sharp:2006zr}. For a comprehensive ussion of the sample selection, the reader is directed to the survey sample selection paper by \citet{Bryant:2015bh}. In brief, the SAMI sample is selected from the Galaxy And Mass Assembly project \citep[GAMA;][]{Driver:2009ai} using stellar-mass cut-offs in redshift bins up to $z < 0.12$. The GAMA sample covers broad ranges in stellar mass and environment. SAMI also includes an addition of eight clusters to probe the higher density environments. Our edge-on discs are selected from the GAMA sample that contains about 830 galaxies at the present stage of the survey (June 2015). 

\begin{figure}
\centering
\includegraphics[width = 8.5cm]{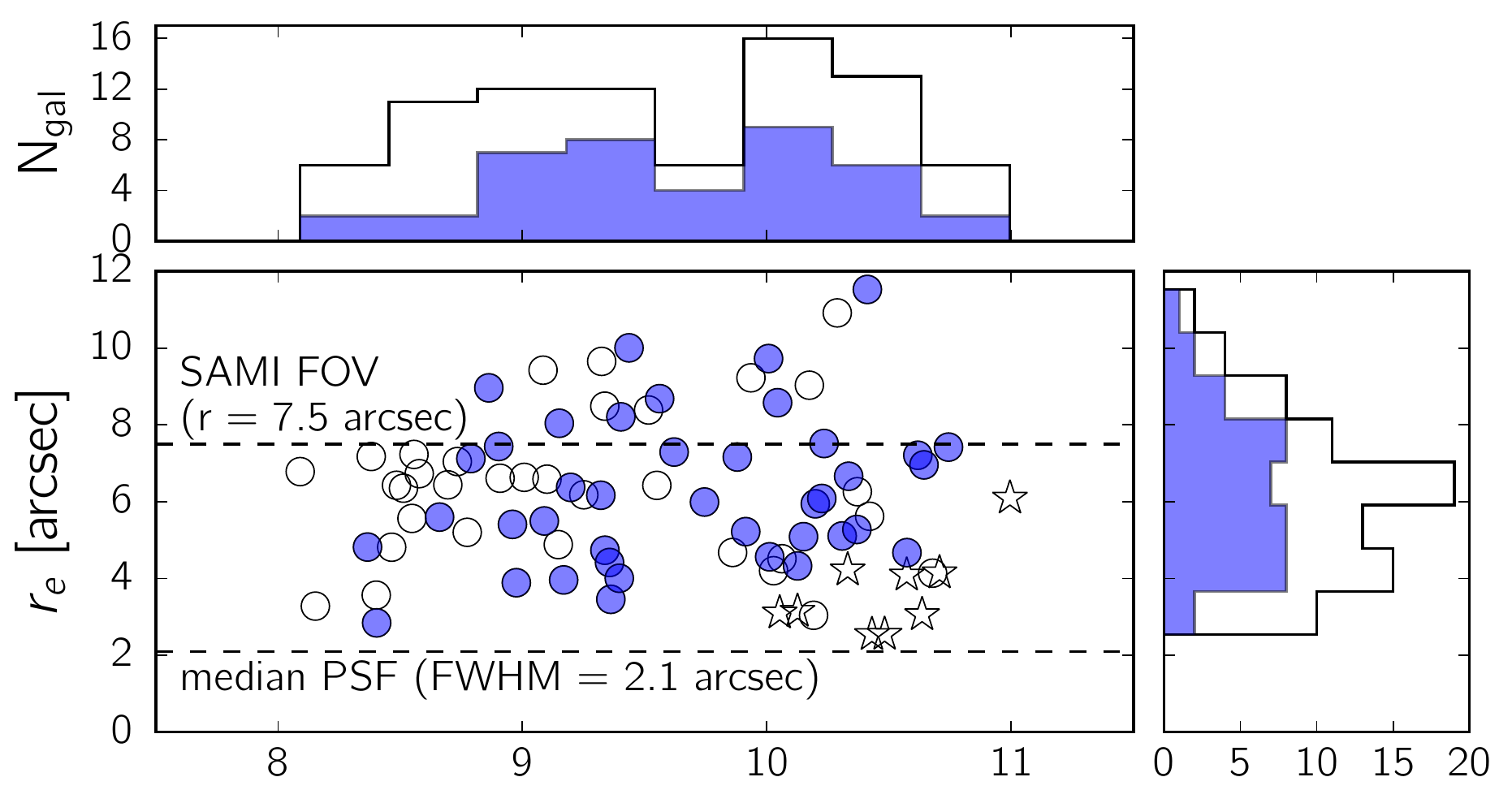}
\includegraphics[width = 8.5cm]{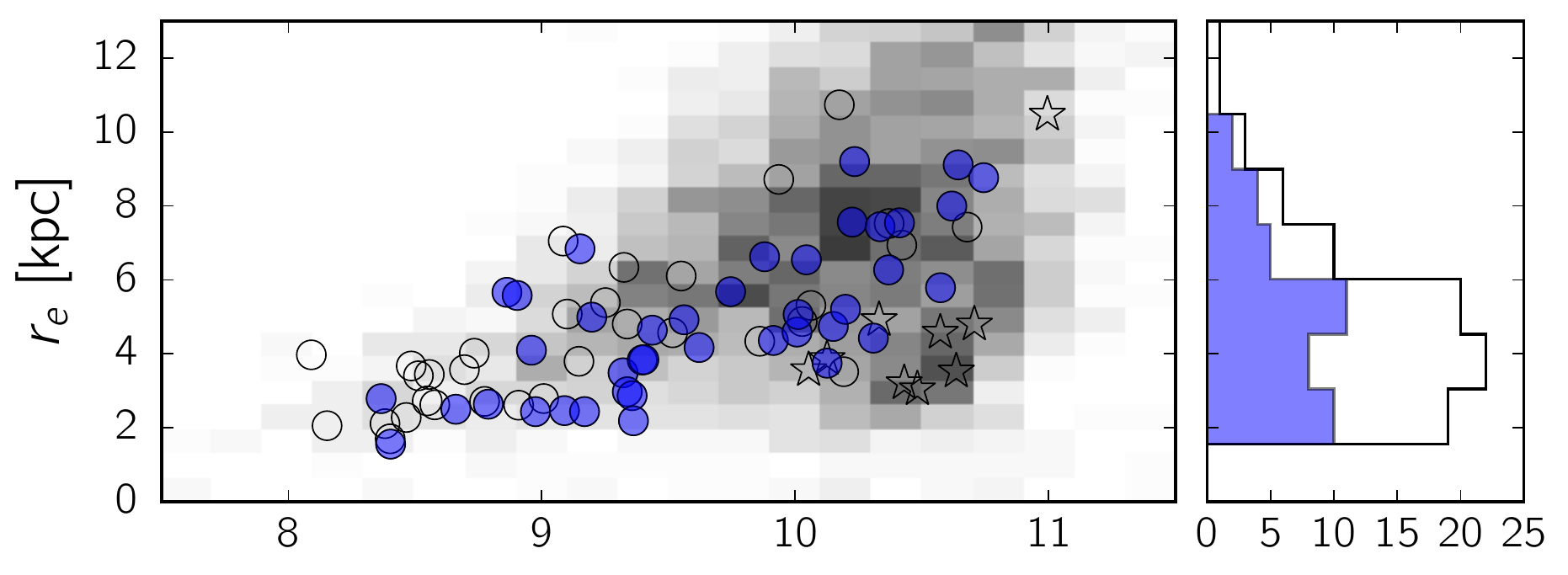}
\includegraphics[width = 8.5cm]{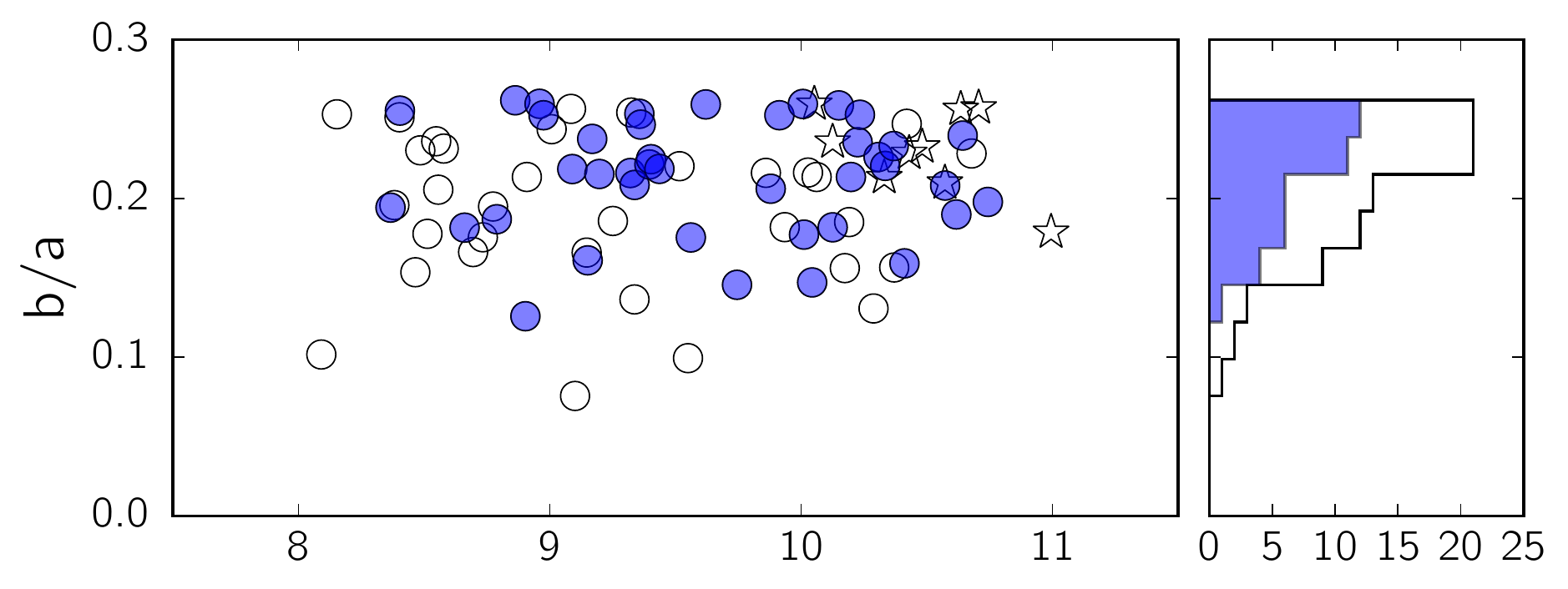}
\includegraphics[width = 8.5cm]{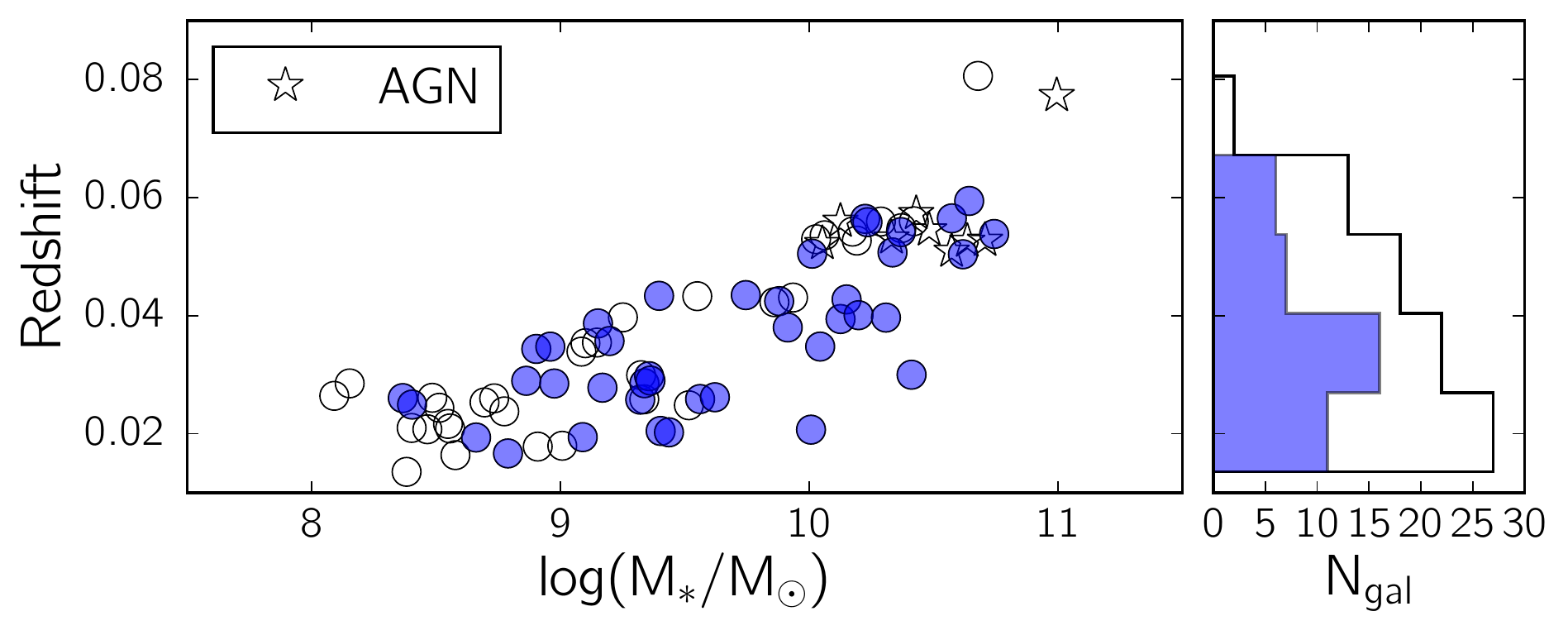}
\vspace{-0.5cm}
\caption{Distributions of our galaxies in stellar mass, {\it r-}band effective radius, minor-to-major axis ratio, and redshift. The circles and stars are the 82 galaxies that satisfy our geometrical (inclination angle $>80$ degrees [{\it b/a} $< 0.26$] and $r_{e}<12$ arcsec) and morphological (not major merger) selections. The open circles are galaxies lacking sufficient H$\alpha$ detections outside 1$r_{e}$ ($\rm N_{gal}=33$) and the open stars are AGNs ($\rm N_{gal}=9$). These galaxies are excluded from this work. Our final sample corresponds to the filled blue circles ($\rm N_{gal}=40$). The blue and black histograms are the distributions for our final sample and all the 82 galaxies, respectively. Grey scale in the $r_{e}$ [kpc] versus stellar mass panel shows the mass-size relation of galaxies in the GAMA survey with inclination angles $>80$ degrees ({\it b/a} $< 0.26$). }\label{fig:distributions}
\end{figure}

\begin{figure*}
\centering
\includegraphics[width = 17cm]{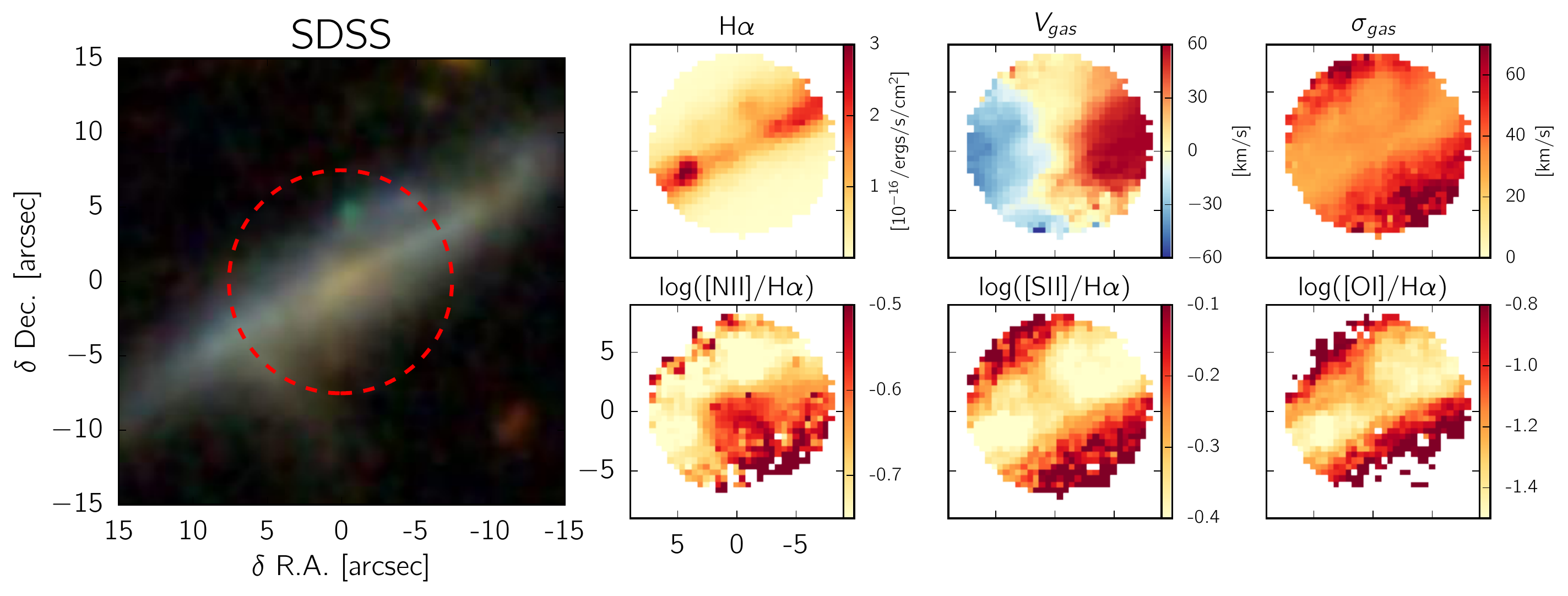}
\caption{Examples of 2D maps reconstructed from the IFS datacubes of GAMAJ115927.23-010918.4 (CATAID: 31452). The dashed red circle in the SDSS 3-colour image (right panel) shows approximately the FOV and pointing of SAMI (15 arcsec in diameter). Note the increase in velocity dispersion and line ratios off the disc plane and the disturbed velocity field are all indicative of galactic winds. We adopt S/N cuts of 3 in these maps. North is up and east is left. }\label{fig:31452}
\end{figure*}

We impose criteria on the galaxy inclination angles and effective radii to ensure that the selected galaxies are edge-on discs well-covered by the SAMI hexabundles. We require minor-to-major axis ratio $(b/a)$ in the {\it r-}band to be less than 0.26, which corresponds to a lower limit on inclination angle of 80 degrees, assuming the classical Hubble formula\footnote{\unexpanded{$\cos^2 i = {(b/a)^2 - q_0^2\over1 - q_0^2}$}, where $i$ is the inclination angle and $q_0$ is a constant of 0.2 ($i=90^\circ$ for $b/a<q_0$).} \citep{Hubble:1926uq}. We also require the {\it r-}band effective radii ($r_{e}$; or half-light radii) to be less than 12 arcsec, such that the SAMI hexabundles cover more than 50\% of the total light for the majority of the sample. The effective radii were measured on Sloan Digital Sky Survey \citep[SDSS;][]{York:2000qe} {\it r-}band images by the GAMA team using {\scshape galfit} \citep{Peng:2010qy,Kelvin:2012rc}. We reject major mergers showing clear tidal features because their complex large-scale winds are not the subject of this study \citep[e.g.][]{Wild:2014fj}. These geometrical and morphological selection criteria yield 82 galaxies. From this sample, we further reject AGNs as those galaxies whose line ratios {[\ion{N}{ii}]~$\lambda$6583}/H$\alpha$ and {[\ion{O}{iii}]~$\lambda$5007}/H$\beta$ at the central spaxels are below the theoretical maximum starburst line by \citet{Kewley:2001lr}. Galaxies lacking enough spaxels outside $1r_{e}$ detected in H$\alpha$ are also not considered because their data do not yield strong constraints on the extraplanar gas properties (see Section~4). These galaxies without good detections on average have lower star formation rates and lower star formation surface densities than those with good detections. After imposing these two criteria, we work with 40 galaxies, representing approximately 5\% of the current SAMI GAMA sample.

Figure~\ref{fig:distributions} shows the stellar mass, redshift, $b/a$, and $r_{e}$ distributions of our sample. The filled circles are our final sample ($\rm N_{gal}=40$). The open symbols (circles and stars) are those satisfying our geometric and morphological selections but either have insufficient detections of extraplanar emissions  ($\rm N_{gal}=33$) or are AGN hosts (open stars; $\rm N_{gal}=9$). We adopt the photometrically-derived stellar masses from the GAMA survey \citep{Taylor:2011kx}. The strong correlation between redshift and stellar mass is a direct result of the selection function of SAMI (see \citealt{Bryant:2015bh} for details). The lack of correlation between $r_e$ in arcsec and stellar mass suggests that our resolution normalized to galaxy size is on average the same for galaxies of different masses .

\section{Data Analysis}

\subsection{Data reduction}
The integral field spectroscopic data are reduced using the automatic data reduction pipeline described in \citet{Sharp:2015ve} and \citet{Allen:2015lq}. We first reduce the spectrum of each fiber following the standard data processing procedures for optical spectroscopy using the {\scshape 2dfdr}\footnote{\href{http://www.aao.gov.au/science/software/2dfdr}{http://www.aao.gov.au/science/software/2dfdr}} data reduction pipeline \citep{Croom:2004fk}, and subsequently resample the spectra on to a rectangular grid of 0.5 arcsec using the drizzle technique \citep{Fruchter:2002lr,Sharp:2015ve}. The reduced data products comprise two datacubes for every galaxy, one for each of the red and blue arms. The blue datacube covers $\approx3750\mbox{--}5800$\AA\ with a spectral sampling of 1.04\AA\ and spectral resolution of $\rm R\approx1750$ or full-width at half-maximum (FWHM) of $\rm\approx170~km~s^{-1}$. The red datacube covers $\approx6300\mbox{--}7425$\AA\ with a spectral sampling of 0.57\AA\ and spectral resolution of $\rm R\approx4500$ or $\rm FWHM\approx65~km~s^{-1}$. The angular resolution of the data is in the range of 1.4 -- 3.0 arcsec, with a median value of 2.1 arcsec. The median angular resolution corresponds to about 0.8 to 2.8 kpc at the distances of our sample. 

\subsection{Emission line fitting}\label{sec-emission-line-fitting}

We apply spectral fitting to extract emission line fluxes and kinematic information from the datacubes. Our emission line fitting toolkit \lzifu\ (Ho et al. in preparation; see also \citealt{Ho:2014uq}) is employed to construct 2-dimensional maps. \lzifu\ adopts the penalized pixel-fitting routine \citep[{\scshape ppxf};][]{Cappellari:2004uq} to model (and subsequently subtract) the stellar continuum and the Levenberg-Marquardt least-squares method \citep{Markwardt:2009lr} to fit Gaussian profiles to the user-assigned emission lines. For fitting the stellar continuum, we  adopt the theoretical simple stellar population (SSP) models assuming Padova isochrones of 24 ages\footnote{equally spaced on a logarithmic scale between 0.004 and 11.220 Gyr} and 3 metallicities\footnote{Z = 0.004, 0.008, and 0.019} from \citet{Gonzalez-Delgado:2005lr}. We model 11 emission lines simultaneously as simple Gaussians requiring them to share the same velocity and velocity dispersion. The 11 emission lines are [\ion{O}{ii}]~$\lambda\lambda$3726,3729, H$\beta$, {[\ion{O}{iii}]~$\lambda\lambda$4959,5007}, {[\ion{O}{i}]~$\lambda$6300}, {[\ion{N}{ii}]~$\lambda\lambda$6548,6583}, H$\alpha$, and {[\ion{S}{ii}]~$\lambda\lambda$6716,6731}. We fix the ratios [\ion{O}{iii}]~$\lambda\lambda$4959/5007 and [\ion{N}{ii}]~$\lambda\lambda$6548/6583 to their theoretical values given by quantum mechanics. Hereafter, we omit the wavelength notation when appropriate, i.e., \OIII $\equiv$ {[\ion{O}{iii}]~$\lambda$5007}, \NII $\equiv$ {[\ion{N}{ii}]~$\lambda$6583}, \OI $\equiv$ {[\ion{O}{i}]~$\lambda$6300}, and \SII $\equiv$ {[\ion{S}{ii}]~$\lambda\lambda$6716,6731}

\begin{figure*}
\centering
\includegraphics[width = 17cm]{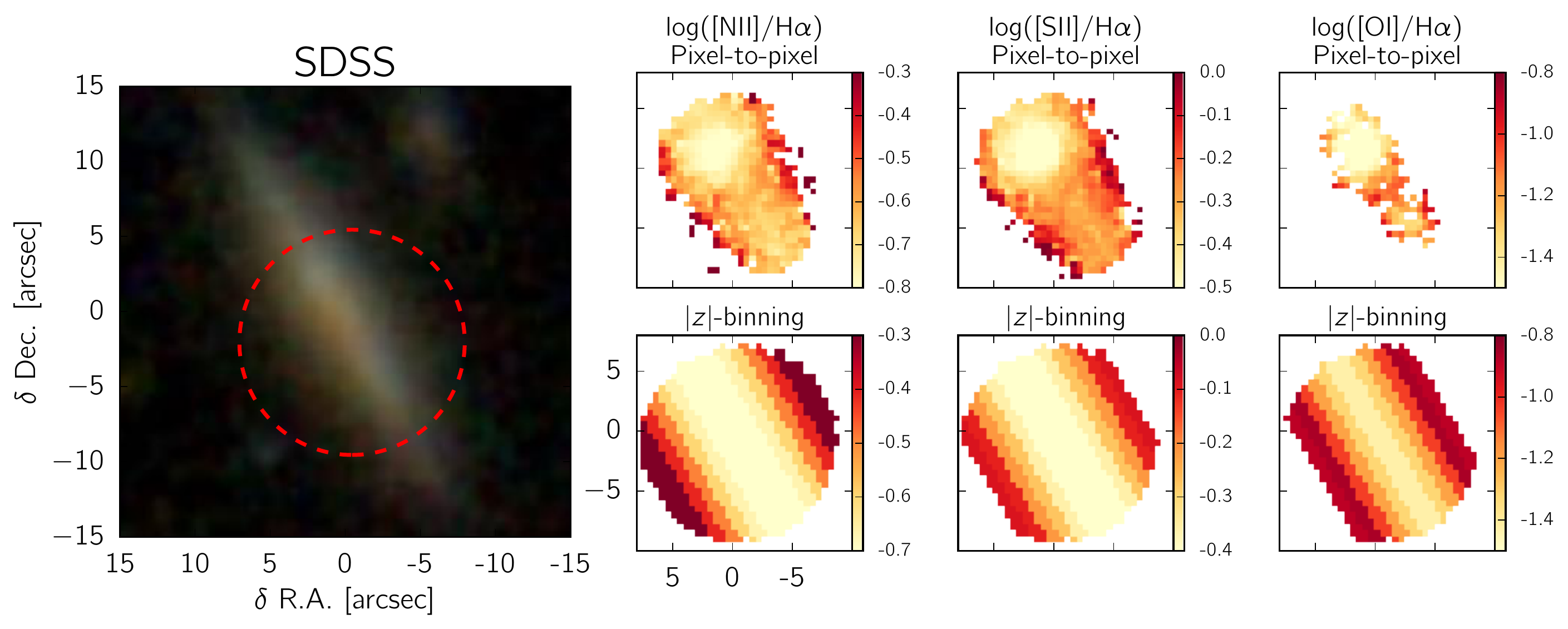}
\caption{Examples of 2D line ratio maps reconstructed from the IFS datacubes of GAMAJ120221.91-012714.1 (CATAID: 185510). The SAMI FOV and pointing position are shown as the red dashed circle in the SDSS 3-colour image (right panel). We create the line ratio maps in two different methods: pixel-to-pixel (upper panels) and adaptive $|z|$-binning (lower panels; see Section~3.2). The adaptive $|z|$-binning bins symmetrically on either side of the disc. The $|z|$-binning maps trace the emission line ratios of the ionized gas unambiguously out to larger $|z|$ than the pixel-to-pixel maps. We adopt S/N cuts of 5 in these maps. }\label{fig:185510}
\end{figure*}

\subsubsection{Pixel-to-pixel maps}

Figure~\ref{fig:31452} shows the emission line flux, line ratio, and kinematic maps of GAMA~J115927.23-010918.4 (CATAID\footnote{CATAID is the galaxy identifier of GAMA}: 31452) and demonstrates the immediate products coming out of our \lzifu\ pipeline. These maps are useful for investigating the properties of the extraplanar gas, as we will elaborate on in later sections. In this particular galaxy, several features are immediately obvious and imply the presence of large-scale galactic winds. The \NII/H$\alpha$, \SII/H$\alpha$ and \OI/H$\alpha$ line ratio maps show that the line ratios increase as a function of height off the disc plane, and suggest that the physical conditions of the ionized gas change dramatically off the disc plane. The enhanced line ratios are likely due to shocks embedded in galactic winds. The gas in partially ionized, post-shock regions emits strong forbidden lines from low ionization species and hydrogen recombination lines \citep{Allen:2008fk}.  The elevated velocity dispersion seen in the $\sigma_{gas}$ map is also consistent with the broad kinematic component seen in IFS data of other wind galaxies \citep[e.g.][]{Rich:2010yq,Rich:2011kx,Fogarty:2012kx,Ho:2014uq}. Two additional interesting features stand out from these maps. In the \NII/H$\alpha$ map, an elevated \NII/H$\alpha$ cone-shaped region pointing south-west could be tracing an ionization cone emerging from the energy injection zone at the centre of the galaxy. We do not observe the other side of the putative bipolar structure presumably due to dust obscuration or intrinsic asymmetry of the winds. This feature is very similar to the bipolar structures that have been observed in the nearby starburst galaxies NGC~253 \citep[][]{Westmoquette:2011uq} and NGC3079 \citep{Cecil:2001uq}. The velocity map of the gas, $v_{gas}$, shows that the overall velocity gradient is mis-aligned with the galaxy disc, which is caused by the complex velocity field of the bipolar winds superimposed on the disc rotation (see Section~4). Although there are no strong indications of on-going interactions in GAMA~J115927.23-010918.4 ($z=0.0202$), the presence of nearby galaxy GAMA~J115923.94-010915.4 (about 50 arcsec or 23 kpc away) at a similar redshift ($z=0.0204$) implies that the environment might play a role in triggering the winds \citep[e.g.][]{Rich:2010yq,Vogt:2013uq}. Higher spatial resolution and multi-wavelength observations are required to determine the precise geometry and cause of the galactic winds in GAMA~J115927.23-010918.4; nevertheless, the pixel-to-pixel maps reconstructed from our IFS data are sufficient to reveal the existence of strong disc-halo interactions through galactic winds.

\subsubsection{Adaptive $|z|$-binning}

GAMA~J115927.23-010918.4 in Figure~\ref{fig:31452} is one of the best cases in our sample where clear wind signatures are detected. However, since  extraplanar line emission is typically much fainter than that of the disc, we spatially bin the datacube parallel to the disc plane ($|z|$-binning). This improves our detection limits at the cost of losing some spatial and kinematic information of the extraplanar gas. 

In the analysis we adopt ``adaptive $|z|$-binning'' to ensure adequate signal-to-noise ratio (S/N) in the following way. First, we produce a bin reference map where the SAMI FOV is divided into 1-arcsec wide slices parallel to the disc (half the typical spatial resolution). We then bin together the required number of slices {\it symmetrically} until the S/N of {[\ion{S}{ii}]~$\lambda$6716 reaches 5. Close to the plane of the disc, the S/N may be already sufficient, leading to a bin containing only the first 1-arcsec-wide slice above and below the disc plane.  As flux drops off at larger distances from the disc plane, we adopt larger bins to obtain the required S/N (i.e. containing several matched slices from above and below the disc).  This symmetric binning assumes that the ionized gas above and below the disc share the same physical properties. Spaxels inside the final bins are summed and the binned spectra are re-analysed by \lzifu.

As a technical note, we take into account non-zero covariance between neighbouring spaxels while binning the datacubes. 
The cubing algorithm in our data reduction pipeline resamples the observed row stacked spectra of multiple dithers on to a rectangular grid, which results in correlated noise between neighbouring spaxels (see \citealt{Sharp:2015ve} and \citealt{Allen:2015lq}). The appropriate way to preserve the variance when binning the data is to bin the datacubes and refit the binned spectra, rather than binning the pixel-to-pixel emission line maps. 

In Figure~\ref{fig:185510}, we show examples of some line ratio maps from fitting the un-binned datacubes and binned datacubes of GAMA~J120221.91-012714.1 (CATAID: 185510). Spaxels or spatial bins with S/N$<5$ on any of the lines associated with the relevant line ratios are masked out. Figure~\ref{fig:185510} demonstrates the advantage of the adaptive $|z|$-binning approach over the traditional pixel-to-pixel approach. The $|z|$-binning maps can trace the ionized gas unambiguously out to larger $|z|$ while single spaxels at such large distances may not survive the S/N cuts and thus provide less of a constraint on the physical conditions of the gas.

\section{Gas kinematics}
While optical line ratios shed light on the underlying excitation sources of the extraplanar gas, the velocity and velocity dispersion of the gas provide additional constraints on the form and strength of disc-halo interactions. 

\begin{figure}
\centering
\includegraphics[width = 8.5cm]{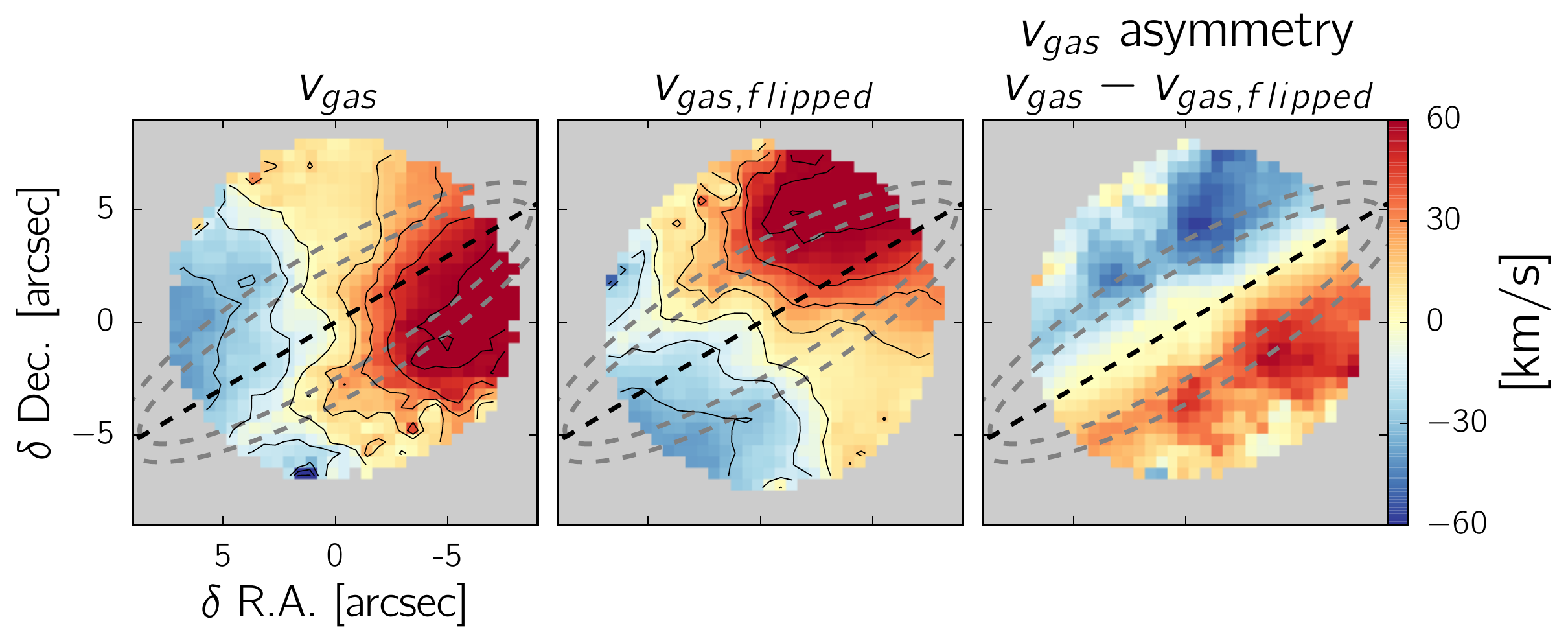}
\vspace{-0.5cm}
\caption{The velocity asymmetry across the major axis of GAMAJ115927.23-010918.4 (CATAID: 31452 ; Figure~\ref{fig:31452}). The middle panel ($v_{gas,flipped}$) is the velocity field ($v_{gas}$) in the left panel flipped over the galaxy major axis. The difference between the left and middle panels is shown in the right panel ($v_{gas}-v_{gas,flipped}$, or the velocity residual map). Contours are equally spaced in 15\kms\ intervals. The dashed black line indicates the major axis of the galaxy. The grey dashed circles indicate concentric ellipses of $1r_e$ (inner) and $1\tilde{r}_e$ (outer) centered at the galaxy optical centre. The outer ellipses are approximately 1 arcsec larger than the inner ellipses. The strong residuals outside $1\tilde{r}_e$ are clear signs of galactic winds. }\label{fig:31452_residual}
\end{figure}

\subsection{Asymmetry of the extraplanar velocity field}
We empirically quantify the strength of disc-halo interactions by the regularity and (a)symmetry of the velocity field. In the cases of weak (or no) disc-halo interactions (e.g. eDIG with no gas flows), the extraplanar gas co-rotates with the disc and usually presents a slight lag in the azimuthal velocity. A vertical gradient in azimuthal velocity of approximately 10 to 30 \kms~kpc$^{-1}$ has been found in nearby galaxies where high quality optical or \HI\ data are available \citep[e.g.][]{Fraternali:2005rr,Heald:2006gf,Heald:2006ul,Heald:2007ve,Zschaechner:2015ff}. There is no strong evidence of pronounced asymmetry in the vertical gradient, i.e. the gas above and below the disc  slows down equally with $|z|$; however, data of high enough quality to constrain the levels of asymmetry remain scarce. In the case of strong disc-halo interactions through galactic winds, the bipolar outflows commonly present opposite velocity fields on either side of the disc. The flow axis is approximately perpendicular to the disc plane but even in extreme edge-on systems the outflow velocity fields are never perfectly symmetric across the major axes of the hosts \citep[e.g.][]{Shopbell:1998yq,Cecil:2001uq,Sharp:2010qy,Westmoquette:2011uq}. The asymmetry of the velocity fields is presumably due to a combination of the outflow axis not lying exactly on the plane of the sky, the clumpy, dusty interstellar medium, the highly structural nature of the wind filaments, and the wide opening angle of the bipolar cones (typically 40 to 60 degrees). 

\begin{figure}
\centering
\includegraphics[width = 8.5cm]{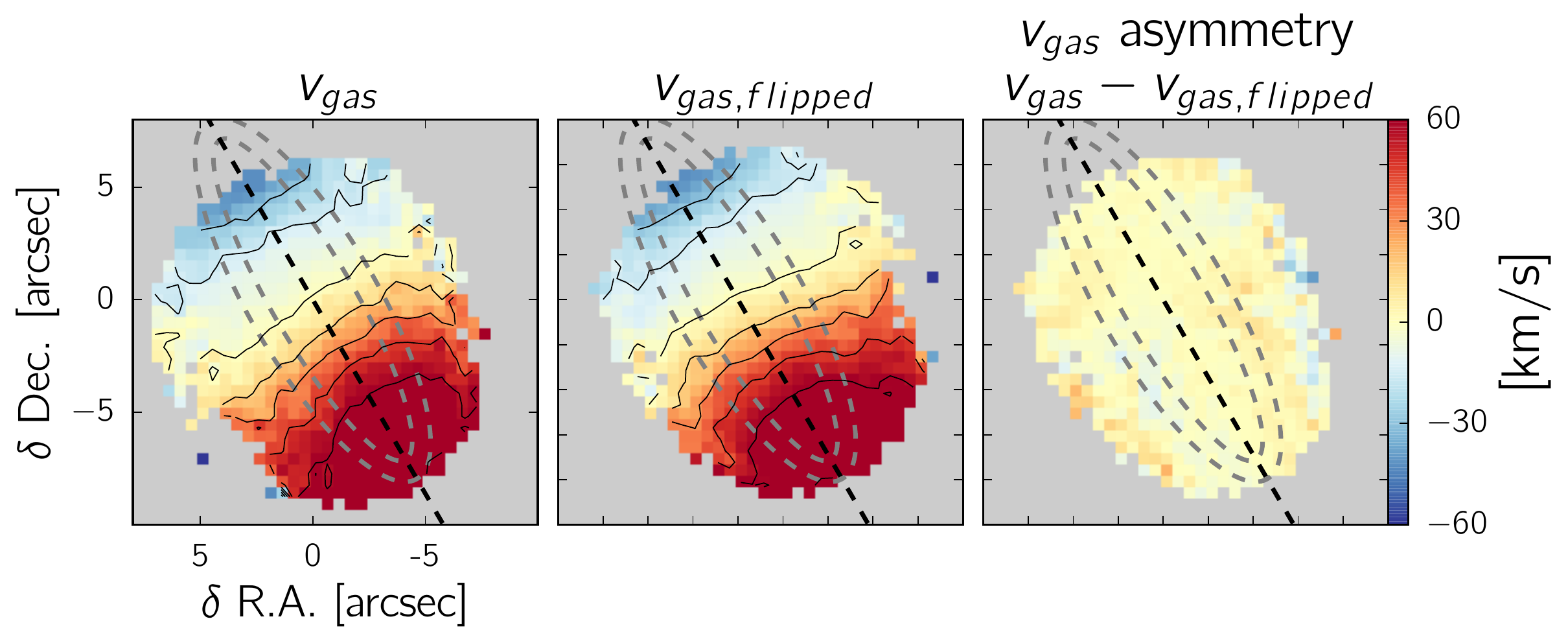}
\vspace{-0.5cm}
\caption{Same as Figure~\ref{fig:31452_residual} but for GAMAJ120221.91-012714.1 (CATAID: 185510 ; Figure~\ref{fig:185510}). The small residuals in the right panel and the line ratio gradients in Figure~\ref{fig:185510} imply the presence of eDIG. }\label{fig:185510_residual}
\end{figure}

Thus, to determine whether the extraplanar emissions trace winds or eDIG, we quantify the velocity asymmetry across the galaxy major axis on the velocity map. We first flip the pixel-to-pixel velocity map over the galaxy major axis, $v_{gas,flipped}$, and subtract the flipped map from the original map (Figures~\ref{fig:31452_residual} and \ref{fig:185510_residual}). We define an ``asymmetry parameter'', $\xi$, that quantifies the flatness of $v_{gas} - v_{gas,flipped}$ (i.e. the velocity asymmetry map):
\begin{equation}\label{eq-xi1}
\xi = {\tilde{\xi}_{+}+\tilde{\xi}_{-} \over 2},
\end{equation}
where 
\begin{equation}\label{eq-xi2}
\tilde{\xi}_{+/-} = \underset{{r_{+/-} > \tilde{r}_e}}{\operatorname{std}} \Big( {v_{gas} - v_{gas,flipped} \over \sqrt{Err(v_{gas})^2 + Err({v}_{gas,flipped})^2}} \Big).
\end{equation}
Here, $Err(v_{gas})$ is the $1\sigma$ map from \lzifu, and $Err({v}_{gas,flipped})$ is the corresponding map for $v_{gas,flipped}$. 
The $\xi$ parameter measures the standard deviation of the S/N of the velocity asymmetry map (i.e. residuals) using only pixels outside the elliptical aperture of $\tilde{r}_e$\footnote{We define the optical centre in our IFS data using the continuum image summed over the red datacube. We fit a S\'{e}rsic profile using {\scshape galfit} to the IFS continuum image while fixing the shape parameters to those determined from the GAMA team fit to the SDSS {\it r-}band image.}. This characteristic radius $\tilde{r}_e$ is the {\it r}-band effective radius increased by approximately 1 arcsec to reduce the effect of beam smearing. Only spaxels with S/N(H$\alpha)>5$ are considered, and all 40 of our galaxies have more than 150 spaxels outside $\tilde{r}_e$. We adopt the mean value of the standard deviations on either side of the disc (i.e. $\tilde{\xi}_{+}$ and $\tilde{\xi}_{-}$) as $\xi$\footnote{We note that $\tilde{\xi}_{+}$ and $\tilde{\xi}_{-}$ in principle contain redundant information but due to interpolation and finite grid size the two values are not exactly the same. We take the mean value to minimise such an artefact.}. As a technical note, to obtain the flipped maps, we first reflect the pixel centres over the galaxy major axis and then linearly interpolate the $v_{gas}$ and $Err(v_{gas})$ maps around the reflected positions.

A perfectly symmetric velocity field would yield $\xi$ of roughly unity as a result of random noise fluctuation in the velocity asymmetry map. The asymmetry parameter $\xi$ becomes much larger than unity when the velocity asymmetry map is not flat due to asymmetry in the velocity field.

Figure~\ref{fig:31452_residual} illustrates our methodology applied to the galaxy in Figure~\ref{fig:31452}, which has the largest $\xi$ in our sample ($\xi = 11.9$). The strong residuals outside $1\tilde{r}_e$ are clear evidence of galactic winds, consistent with the line ratio, structural, and kinematic signatures seen in Figure~\ref{fig:31452}. The flat residuals inside $1\tilde{r}_e$ indicate the good symmetry of the prominent disc rotation. Figure~\ref{fig:185510_residual} shows the same application on the galaxy in Figure~\ref{fig:185510}. No strong residuals are visible on the velocity asymmetry map ($v_{gas} - v_{gas,flipped}$; $\xi = 1.3$) despite the clear line ratio gradients seen in Figure~\ref{fig:185510}, implying the presence of eDIG rather than shocked outflowing gas. These sets of figures demonstrate the importance of using both kinematics and line ratios to investigate galactic winds and eDIG .

The parameter $\xi$ can be used to trace winds or eDIG under the assumptions that our edge-on discs are regular and not heavily warped. Mergers are excluded from our sample (Section~2), and all our galaxies present regular rotation on their discs, consistent with not undergoing major interactions (see below Figures~\ref{fig:wind0} and \ref{fig:nowind0}). The SDSS images also suggest that all the galaxies have regular discs without strong warp signatures. Although warping of stellar discs is common, the warps are only important at large radii (3--6 times the disc scalelength or 2--4$r_e$; \citealt{Saha:2009ly}), typically beyond the SAMI FOV for the majority of our sample (90\%). Thus, our velocity asymmetry parameter predominately reflects how the extraplanar gas is effected by winds.

It is worth pointing out that our approach of quantifying velocity asymmetry does not depend on any kinematic modeling, which can be non-trivial due to the edge-on viewing angle. Future work should develop 3-dimensional models that include the effects of beam smearing (such as the approach by \citealt{Jozsa:2007ul} and \citealt{Di-Teodoro:2015cl}) and radiative transfer to reproduce the emission line data cubes.

\begin{figure*}
\includegraphics[width = 18cm]{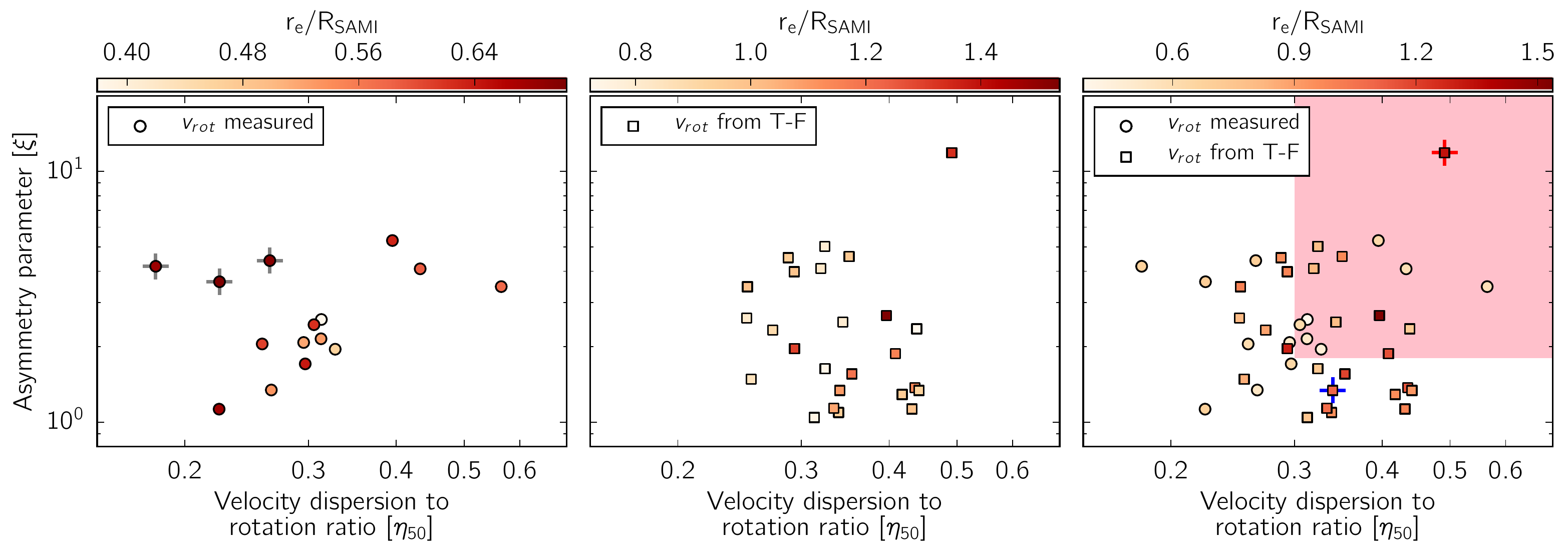}
\vspace{-0.5cm}
\caption{The asymmetry parameter of the velocity field $\xi$ versus the velocity dispersion to rotation ratio parameter $\eta_{50}$ of our sample. The left panel presents galaxies with $v_{rot}$ measured from the SAMI data,  the middle panel presents galaxies with $v_{rot}$ inferred from the Tully-Fisher relation, and the right panel shows all the galaxies. The three high-$\xi$ low-$\eta_{50}$ galaxies marked by the grey crosses are ussed in Section~4.3. The red cross in the right panel corresponds to the example galaxy shown in Figures~\ref{fig:31452} and \ref{fig:31452_residual}; and the blue cross in Figures~\ref{fig:185510} and \ref{fig:185510_residual}. Galaxies falling in the shaded region in the right panel are identified as wind-dominated galaxies (see Figure~\ref{fig:wind0}). }\label{fig:xi-eta}
\end{figure*}

\subsection{Elevated extraplanar velocity dispersion}
We also make use of the velocity dispersion of the extraplanar gas to assess the form and strength of disc-halo interactions. In the case of weak (or no) interactions, the velocity dispersion of eDIG arises predominately from line-of-sight projection of the gas corotation with the disc. \citet{Heald:2006gf,Heald:2006ul,Heald:2007ve} found extraplanar $\sigma_{gas}\lesssim50$~\kms\ in NGC~891, NGC~5775, and NGC~4302. These three edge-on discs have very similar ionized gas rotation velocities ($v_{rot}\approx 170\mbox{--}200$~\kms) and stellar masses ($\log(\rm M_*/M_{\sun})\approx10.7$). Their findings imply that $\sigma_{gas}/v_{rot}\lesssim0.3$ for eDIG. In the case of strong interactions, the velocity dispersion of the extraplanar gas is broadened by both the turbulent motion of the outflowing gas and line splitting caused by emissions from the approaching and receding sides of the outflow cones. The magnitude of the line splitting depends on outflow velocities, topology and inclination angle. In classical wind galaxies, line splitting of about 100~\kms\ is common \citep[e.g.][]{Heckman:1990lr,Sharp:2010qy}. In the extreme case of M82, line splitting as high as 300~\kms can be easily resolved by optical spectrographs \citep{Axon:1978fv,Shopbell:1998yq}. A very broad component (FWHM of 300~\kms) from the halo gas has also been observed in M82 \citep[][]{Bland:1988yq}. In less extreme winds, the line splitting could be too narrow for SAMI to resolve, especially if viewing the outflow cones from the side where the projected velocities are small. In fact, we do not see clear indications of line splitting in the majority of our wind candidates given the S/N of our data. Nevertheless, the increase in velocity dispersion as a result of unresolved line splitting is still strong evidence of the presence of outflows. 

To quantify the extraplanar velocity dispersion, we invoke a simple parameter $\eta_{50}$, the ``velocity dispersion to rotation ratio'' parameter, defined as
\begin{equation}
\eta_{50} = \sigma_{50} / v_{rot}, 
\end{equation}
where $\sigma_{50}$ is the median velocity dispersion of all spaxels outside $1\tilde{r}_e$ with S/N(H$\alpha)>5$. As before, all 40 of our galaxies have more than 150 spaxels meeting our S/N criterion outside $1\tilde{r}_e$. The rotation velocity is measured on our velocity map from the spaxels along the optical major axis with the maximum velocity. This method is applied only when our FOV extends to $1.4r_e$ ($<R_{\rm SAMI} \approx7.5$ arcsec). For an exponential disc, this radial distance corresponds to about 2.4 times the disc scale length at which typical rotation curves already reach their maximum velocities \citep{Sofue:2001kl,Cecil:2015rz}. For galaxies without sufficiently large spatial coverage, we use the stellar mass Tully-Fisher relation to infer $v_{rot}$ \citep{Bell:2001yq}. We note that for our purpose $v_{rot}$ should ideally be the maximum rotation velocity of the ionized gas, and we approximate it using the Tully-Fisher relation. The stellar mass Tully-Fisher relation unfortunately exhibits a scatter of about 0.5~dex, which directly translates to a factor of 3 in the systematic uncertainty of $\eta_{50}$.

\subsection{An empirical identification of wind-dominated galaxies}

Based on the two different empirical parameters for quantifying the strength of disc-halo interactions, it should be possible to distinguish galactic winds from eDIG. Galactic winds are expected to show {\it both} high $\xi$ and $\eta_{50}$, whereas eDIG has low $\xi$ and $\eta_{50}$. That is, galactic winds both disturb the symmetry of the extraplanar velocity field and increase the extraplanar emission line widths, whereas eDIG should still follow the velocity field of the galaxy. If winds are the only mechanism disrupting the extraplanar gas, then a trend between $\xi$ and $\eta_{50}$ is expected. Figure~\ref{fig:xi-eta} shows our sample in the $\xi$ versus $\eta_{50}$ parameter space. We present separately galaxies with $v_{rot}$ measured (left panel) and $v_{rot}$ inferred from the Tully-Fisher relation (middle panel). 

For galaxies with directly measured $v_{rot}$, it is inconclusive whether there is a trend between $\xi$ and $\eta_{50}$. There are no galaxies in the lower right corner ($\eta_{50} \gtrsim 0.4$ and $\xi \lesssim 3$), however, there are three galaxies (marked with crosses) with larger $\xi$ than the other galaxies of similar $\eta_{50}$. The Spearman rank correlation test implies no significant correlation for the galaxies in the left panel ($\rho$ of 0.21 and {\it p-}value of 0.46), unless the three high-$\xi$ low-$\eta_{50}$ galaxies are excluded from the test ($\rho$ of 0.80 and {\it p-}value of 0.0016). We speculate that there might be additional mechanisms responsible for causing the asymmetry of the extraplanar gas (see Section~7.3). A trend between $\xi$ and $\eta_{50}$ can not be confirmed by our medium-sized sample.

\begin{figure*}
\includegraphics[width = 18cm]{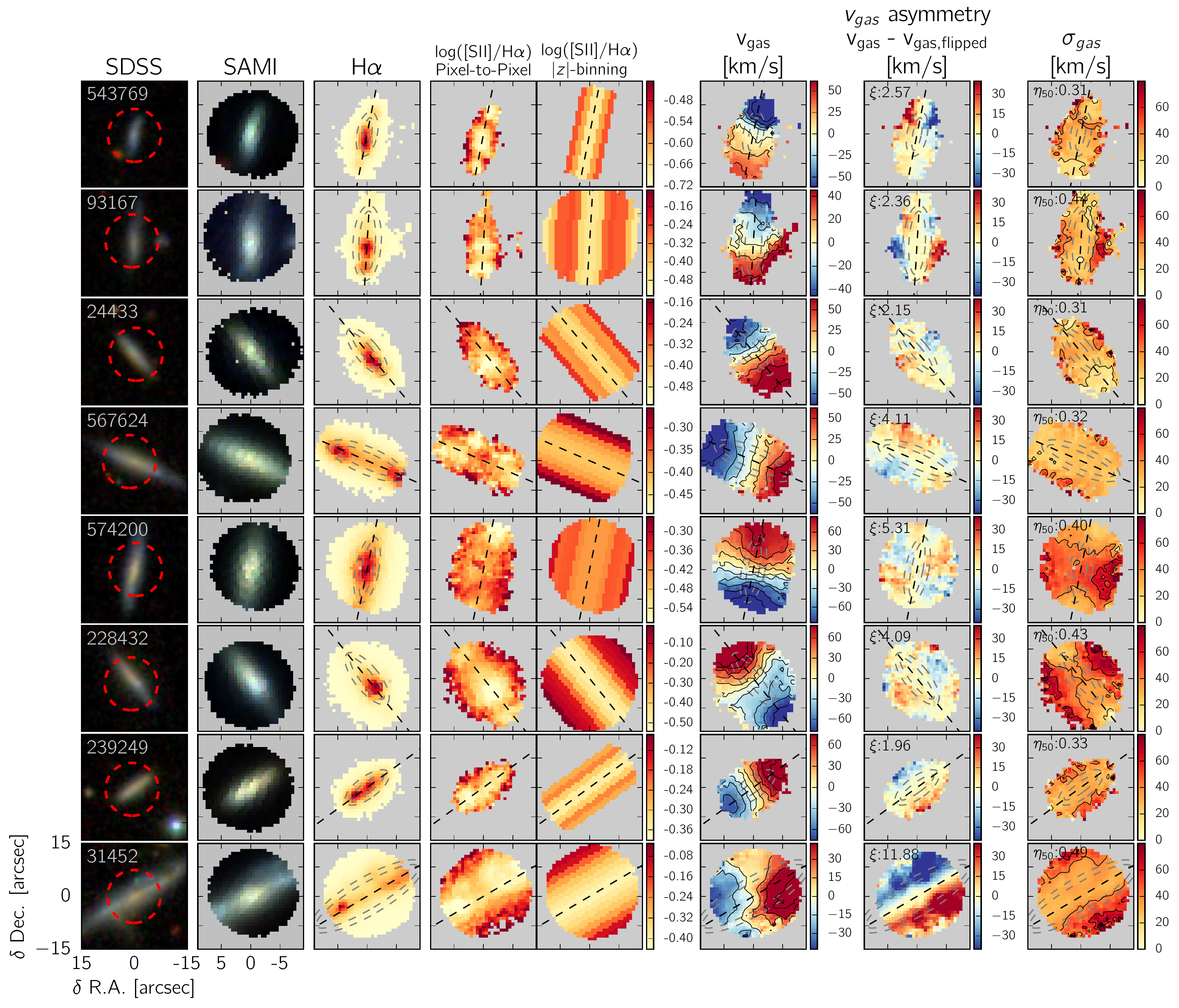}
\vspace{-0.5cm}
\caption{From left to right: SDSS 3-colour, SAMI 3-colour (red: 6300\AA; green: 4800\AA; blue: 4000\AA), H$\alpha$, \SII/H$\alpha$ (pixel-to-pixel and adaptive $|z|$-binning), velocity, velocity asymmetry, and velocity dispersion maps of our wind-dominated galaxies ($\eta_{50} > 0.3$ and $\xi > 1.8$). The adaptive $|z|$-binning maps present the average \SII/H$\alpha$ measured in different bins (Section~3.2.2). GAMA CATAIDs and the $\eta_{50}$ and $\xi$ values are labelled in the plots. The velocity and velocity dispersion contours are equally spaced in 20\kms\ intervals. The red dashed circles on the SDSS images indicate approximately the FOV and pointings of the SAMI hexabundles. The dashed lines indicate the galaxy major axes. The inner and outer concentric ellipses correspond to elliptical apertures of $1r_e$ and $1\tilde{r}_e$, respectively. The outer ellipses are approximately 1 arcsec larger than the inner ellipses. }\label{fig:wind0}
\end{figure*}

For galaxies with $v_{rot}$ inferred from the Tully-Fisher relation, there is also no clear relationship between $\xi$ and $\eta_{50}$ and the Spearman rank correlation test suggests no significant correlation ($\rho$ of $-0.19$ and {\it p-}value of 0.35). The lack of correlation is expected given that the scatter of Tully-Fisher relation (0.5~dex) yields a factor of 3 in the systematic uncertainty of $\eta_{50}$.

\begin{figure*}
\includegraphics[width = 18cm]{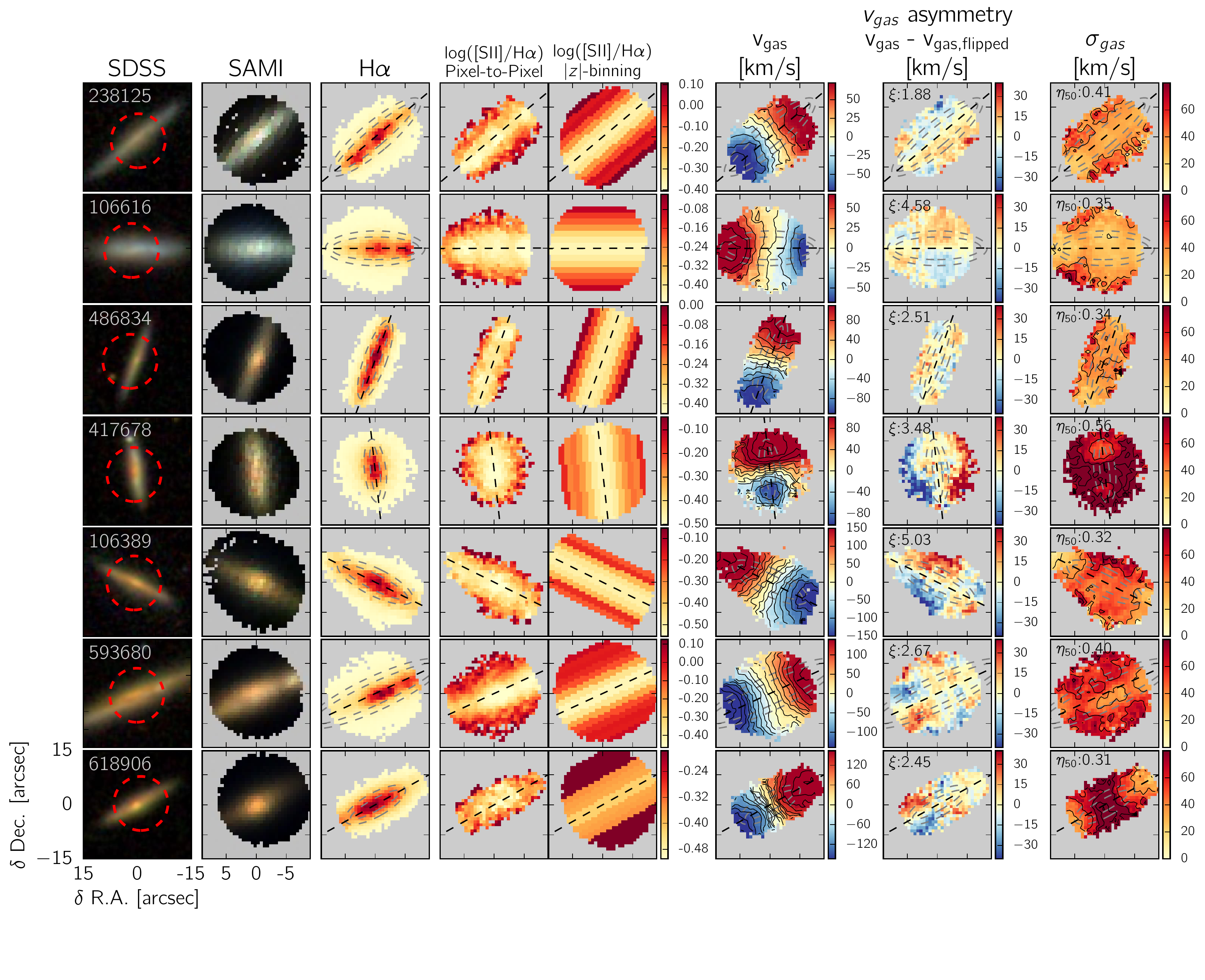}
\vspace{-1.3cm}
\contcaption{}\label{}
\end{figure*}

 We conservatively select galaxies whose extraplanar gas is affected by galactic winds from the upper right corner of Figure~\ref{fig:xi-eta}. We select galaxies with $\eta_{50} > 0.3 $ and $\xi >1.8 $, as defined by the shaded region in the right panel. These criteria yield 15 galaxies. For the purpose of presentation, we will henceforth refer these galaxies as ``wind-dominated galaxies''. Figure~\ref{fig:wind0}  presents their velocity, velocity asymmetry, velocity dispersion, and \SII/H$\alpha$ line ratio maps. The remaining galaxies are presented in the appendix (Figure~\ref{fig:nowind0}). We note that it is, to some extent, unphysical to impose hard limits to identify wind galaxies. The distribution of our galaxies in the $\xi$ versus $\eta_{50}$ parameter space does not favour the idea that there exists distinct groups of galaxies with similar kinematic properties of their extraplanar gas; instead, the distribution could imply a continuous ``strength sequence'' of disc-halo interaction, connecting the very quiescent, non-interacting (or weakly interacting) eDIG to the very violent, strongly interacting classical galactic winds. It is likely that in galaxies not classified as wind-dominated, there could still be localized feedback from \HII\ regions driving gas away from the discs, disturbing the extraplanar gas kinematics at small scales, and causing a slight increase in  $\xi$ and/or $\eta_{50}$. In fact, some galaxies (e.g. CATAID: 496966, 228066, 348116, 517594, 595027; Figure~\ref{fig:nowind0}) indeed show indications of localised velocity asymmetry possibly driven by off-centre starbursts, consistent with the galactic chimney picture \citep{Norman:1989pi}. It is also likely that in some of the wind-dominated galaxies, a portion of the pre-existing eDIG remains co-rotating with the discs while the outflowing gas interacts with the rest of the eDIG. Given our sample size, it is nonetheless convenient to invoke a binary classification scheme in order to investigate the underlying physical processes governing the strength of disc-halo interactions.

The wind-dominated galaxies in Figure~\ref{fig:wind0}, by selection, show strong velocity asymmetry and elevated extraplanar velocity dispersion. Classical galactic wind signatures can be easily identified in these maps. Some galaxies (e.g. CATAID: 228432, 574200) present ionization cone signatures on their velocity dispersion maps and their velocity asymmetry maps, with strong velocity residuals tracing the limbs of the putative bipolar cones. In others (e.g.  CATAID: 31452, 239249, 417678, 106389), the velocity asymmetry maps correlate spatially with the velocity dispersion maps and usually present positive/negative residuals on either sides of the discs. On the contrary, the remaining galaxies presented in the appendix do not show these kinematic signatures of galactic winds as clearly as the wind-dominated galaxies. In all the wind-dominated galaxies, the \SII/H$\alpha$ line ratio increases with $|z|$. This is partly due to excitation by shocks embedded in the outflows (see Section~6.2). 

\section{Host galaxy properties}
We compute the host galaxy properties before returning to investigate the underlying processes driving the different properties of the extraplanar gas. The host galaxy properties we investigate are stellar mass, star formation rate, star formation history, and their associated quantities. As before, we adopt the photometrically-derived stellar mass from the GAMA survey \citep{Taylor:2011kx}. The following two subsections describe how we measure SFRs using spectral energy distribution (SED) fitting and H$\alpha$ flux, and star formation histories using the \Dn\ and \Hd\ indices.

\subsection{Star formation rate}
To estimate the SFR, we adopt the SED fitting code \magphys\ described in detail in \citet{da-Cunha:2008fk}. \magphys\ is a Bayesian-based package that uses observed photometry to constrain stellar population synthesis models by \citet{Bruzual:2003qy} and dust emission of different temperatures from molecular clouds and the diffuse ISM. The UV-to-IR SEDs assume exponential star formation histories of different star formation time-scales with random bursts of star formation superimposed. Physical parameters of the models (SFR, star formation history, stellar mass, dust temperature, etc.) and corresponding errors are constrained by comparing synthesized and observed photometry under a Bayesian framework. We utilize photometry from the Galaxy Evolution Explorer (GALEX; FUV, NUV; \citealt{Martin:2005ul,Wyder:2005pd}), SDSS ({\it u,g,r,i,z}; \citealt{Adelman-McCarthy:2008vn}), UKIRT Infrared Deep Sky Survey (UKIDSS) Large Area Survey (Y, J, H, K; \citealt{Lawrence:2007dq}) and Herschel Space Observatory (Photodetecting Array Camera and Spectrometer: 100$\micron$ and 160$\micron$; Spectral and Photometric Imaging Receiver or SPIRE: 250$\micron$, 350$\micron$, and 500$\micron$; \citealt{Pilbratt:2010eu,Poglitsch:2010vn,Griffin:2010jk}). The SDSS and UKIDSS photometry are {\it r}-band aperture-matched by the GAMA team, and are compiled in their data release 2 \citep{Hill:2011qf,Liske:2015rf}. The Herschel photometry comes from the Herschel-ATLAS survey \citep{Eales:2010uq} Phase 1 Internal Data Release. The maps and catalogues will be described in Valiante et al (in preparation) and have been processed in a similar way to that described by \citet{Ibar:2010fj}, \citet{Pascale:2011kx} and \citet{Rigby:2011vn}. The matching to optical data was performed in a similar way to that described in \citet{Smith:2011yq}, and will be presented in Bourne et al. (in preparation). The GALEX, SDSS, and UKIDSS photometry are available for almost all galaxies: 38/40, 40/40, and 39/40, respectively. More than half of our sample (28/40) has photometry from the Herschel-ATLAS survey. Of the 12 galaxies without Herschel photometry, two of them fall outside the Herschel-ATLAS SPIRE footprints and the remainder falls below the detection limits at 250$\micron$.

In addition to the SED-based SFRs, we also calculate SFRs based on the H$\alpha$ line measured in our integral field data. We construct global, one-dimensional spectra by summing all spaxels in the datacubes, and fit the spectra using the same  method described in Section~3.2. The H$\alpha$ fluxes are extinction corrected using the Balmer decrement method assuming an intrinsic H$\alpha$ to H$\beta$ line ratio of 2.86 under case-B recombination of $T_e$ of 10,000~K and $n_e$ of 100 cm$^{-3}$ \citep{Osterbrock:2006lq}. We adopt the extinction law by \citet{Cardelli:1989qy} of $R_V$ of 3.1. The extinction corrected H$\alpha$ line fluxes are then converted to SFRs following \citet{Murphy:2012fv}. We note that a \citet{Chabrier:2003uq} initial mass function (IMF) is assumed in the \citet{Bruzual:2003qy} models used in \magphys, and a \citet{Kroupa:2001fv} IMF is assumed for the H$\alpha$ SFRs; however the systematic errors due to slight differences between the IMFs are small (within 10\%; \citealt{Bell:2007fr}).

\begin{figure}
\centering
\includegraphics[width = 8.5cm]{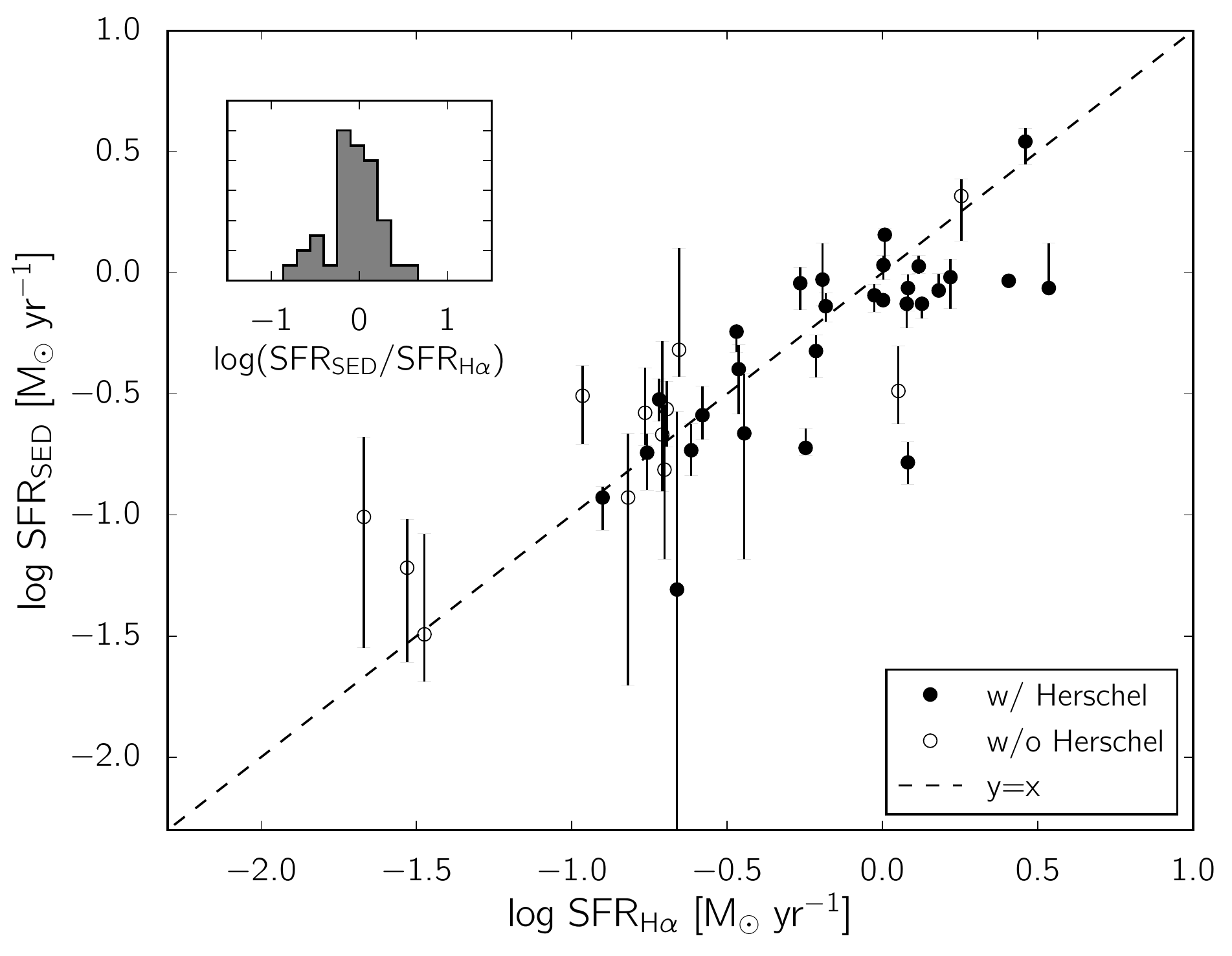}
\vspace{-0.5cm}
\caption{Comparison between SFRs derived from SED fitting using \magphys\ and from global H$\alpha$ fluxes measured in our IFS data. Galaxies with Herschel photometry from the Herschel-ATLAS survey are shown as solid circles and the remainders in open circles. The error bars indicate $\pm1\sigma$ confidence intervals reported by \magphys\ using Bayesian inference. The H$\alpha$ SFR errors are negligible ($\ll10\%$) compared to the ${\rm SFR_{SED}}$ errors and systematic errors due to star formation calibrations. The dashed line is the one-to-one line. The inset shows the distribution of the logarithmic ratios characterised by median $\pm$ standard deviation of $-0.02\pm0.30$~dex. }\label{fig:compare-sfr}
\end{figure}

Figure~\ref{fig:compare-sfr} compares the SED- and H$\alpha$-based SFRs. The presence of a positive correlation without an appreciable offset (approximately $-0.02$~dex) is encouraging since both SFR diagnostics are subject to different systematic uncertainties and the two SFRs probe star formation of different time-scales (H$\alpha$: a few Myr; \magphys: average over the past 100 Myr). The H$\alpha$-based SFRs are subject to uncertainties in extinction correction, which can be particularly important in edge-on discs when viewing through prominent dust lanes. Furthermore, correcting for Balmer absorption relies on accurate stellar population synthesis, which can be problematic when the Balmer line equivalent widths are large (i.e. weak emission, strong absorption). Also, the H$\alpha$ emission could also come from non-thermal ionization by shocks. Finally, although the covering fractions are high (typically $\gtrsim50\%$}), there could still be missing H$\alpha$ fluxes outside the apertures. The SED-based method depends critically on availability of multi-band photometry (particularly in far infrared), and accurate stellar population synthesis models (see the reviews by \citealt{Walcher:2011lr}; \citealt{Conroy:2013lr}). With these caveats, we consider the scatter in Figure~\ref{fig:compare-sfr} as an empirical estimate of the degree of uncertainty in the SFRs (approximately 0.3~dex). We will make use of both types of SFRs and draw qualitative conclusions in a statistical sense.

\subsection{Star formation history: \Dn\ and \Hd}
To quantify the star formation histories, we make use of the classical \Dn\ and \Hd\ indices. Combination of the two indices has been employed as a simple and robust diagnostic to distinguish between continuous and bursty star formation histories \citep[][]{Kauffmann:2003yq}. The \Dn\ index measures the strength of the 4000\AA\ break that monotonically increases with the age of the stellar populations. The H$\delta$ absorption line index, \Hd, is sensitive to bursts of star formation, and increases monotonically until peaking at $\sim0.1\mbox{--}1$ Gyr after a burst of star formation. 

We measure the global \Dn\ and \Hd\ indices on one-dimensional global spectra extracted from the datacubes using elliptical apertures of $1r_e$. The \Dn\ index follows the original definition by \citet{Bruzual-A.:1983rc} but with the narrower passbands proposed by \citet{Balogh:1999fp}. Where possible (in 31 out of 40 galaxies), we calibrate our \Dn\ to SDSS DR7 values by comparing \Dn\ measured with data extracted from the same SDSS apertures with those from the MPA-JHU value-added catalog\footnote{\href{http://mpa-garching.mpg.de/SDSS/DR7/}{http://mpa-garching.mpg.de/SDSS/DR7/}}. This empirical correction is to compensate for potential calibration errors in the blue end of the SAMI spectrograph. The multiplication correction factors are typically close to unity (0.9 -- 1.1). The \Hd\ index measures the absorption equivalent width of H$\delta$ (where positive/negative corresponds absorption/emission) using the passbands defined in the Lick system \citep{Worthey:1997fk}. To remove contamination from the nebular emission, we include the H$\delta$ emission line in our \lzifu\ fit and subtract the flux when the line S/N is greater than 3. One galaxy (CATAID:184237) is excluded from the forthcoming comparisons related to star formation history because its weak weak blue continuum does not yield robust index measurements. 
\begin{figure*}
\centering
\includegraphics[width = 18cm]{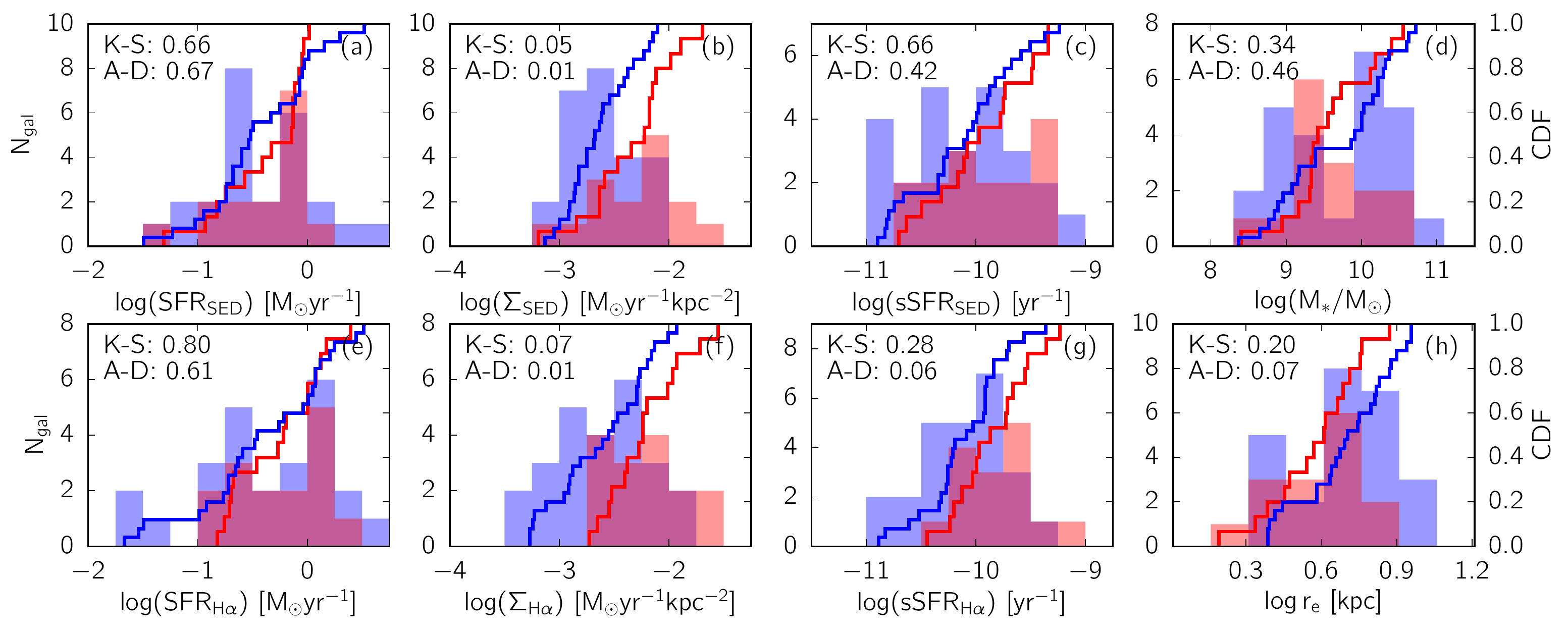}
\vspace{-0.5cm}
\caption{Distributions of the wind-dominated galaxies (red) and the remaining galaxies (blue) in SFR, SFR surface density ($\Sigma$), specific SFR (sSFR), stellar mass ($M_*$) and effective radius ($r_e$). Both histograms (left axes) and cumulative distributions (CDFs; right axes) are shown. Panels (a), (b) and (c) correspond to SFRs derived from SED fitting, and panels (e), (f) and (g) correspond to those from H$\alpha$. The {\it p-}values of two-sample Kolmogorov-Smirnov test and Anderson-Darling test are labelled in each panel.  }\label{fig:corr_global}
\end{figure*}

\section{Results}

\subsection{Correlation with global quantities}

\subsubsection{SFR, SFR surface density, specific SFR, stellar mass and size}
In Figure~\ref{fig:corr_global}, we present the logarithmic distributions of SFR, SFR surface density ($\Sigma\equiv {\rm SFR}/2\pi r_e^2$), specific SFR ($\rm sSFR\equiv SFR/M_*$), stellar mass, and effective radius. We show both SFRs derived from fitting the SED (panels (a), (b) and (c)) and H$\alpha$ (panels (e), (f) and (g)). The stellar mass and size distributions are shown in panels (d) and (h), respectively. We colour the identified wind-dominated galaxies in red and the remaining galaxies in blue. To test if the two groups of galaxies have similar properties, we perform two-sample Kolmogorov-Smirnov (K-S) tests and Anderson-Darling (A-D) tests, and  label the {\it p}-values in each panel. A smaller {\it p}-value indicates that the two distributions are more likely drawn from different parent populations. Typically, a {\it p}-value $<0.05$ is considered statistically significant and {\it p}-value $<0.01$ highly significant \citep[e.g.][]{Taylor:1997qy}. The A-D test is more robust than the K-S test when the differences between cumulative distribution curves are most prominent near the beginning or end of the distributions.

Figure~\ref{fig:corr_global} shows that the identified wind-dominated galaxies span wide ranges in SFR, SFR surface density, sSFR, and stellar mass. The wind-dominated galaxies have SFRs as low as approximately $0.1~\rm M_{\sun}~yr^{-1}$, and SFR surface densities of approximately $10^{-3}$ to $10^{-1.5}\rm ~M_{\sun}~yr^{-1}~kpc^{-2}$, much lower than the canonical threshold found in classical wind galaxies \citep[$\Sigma> 0.1~\rm M_{\sun}~yr^{-1}~kpc^{-2}$;][]{Heckman:2002fk}.

Comparing the distributions of the two groups of galaxies, SFR surface density is the only parameter that presents significantly different distributions for the two groups of galaxies. The wind-dominated galaxies on average have higher SFR surface densities. The same conclusion can be drawn from both the SED- and H$\alpha$-based SFRs. The {\it p-}values are 0.01 for the A-D tests, and 0.05 and 0.07 for the K-S tests. There is no evidence that the SFR distributions are drawn from different parent populations, indicating that global SFRs do not determine wind activities, but rather that the critical parameter is the spatial concentration of star formation (Section~7.1). The differences in SFR surface density may be partially driven by the differences in size (panel h); the A-D {\it p-}value of 0.07 is only marginally higher than the threshold for  significance. The only other {\it p-}value implying a possible significant difference is from the A-D test of the H$\alpha$ sSFRs ({\it p-}value of 0.06). This may be suggestive of a stronger correlation between winds and more recent star forming activities (a few Myr, as traced by H$\alpha$) than the star forming activities on longer time-scales (100 Myr, as traced by {\scshape magphys}).

\subsubsection{Galaxy main sequence}

\begin{figure*}
\centering
\includegraphics[width = 18cm]{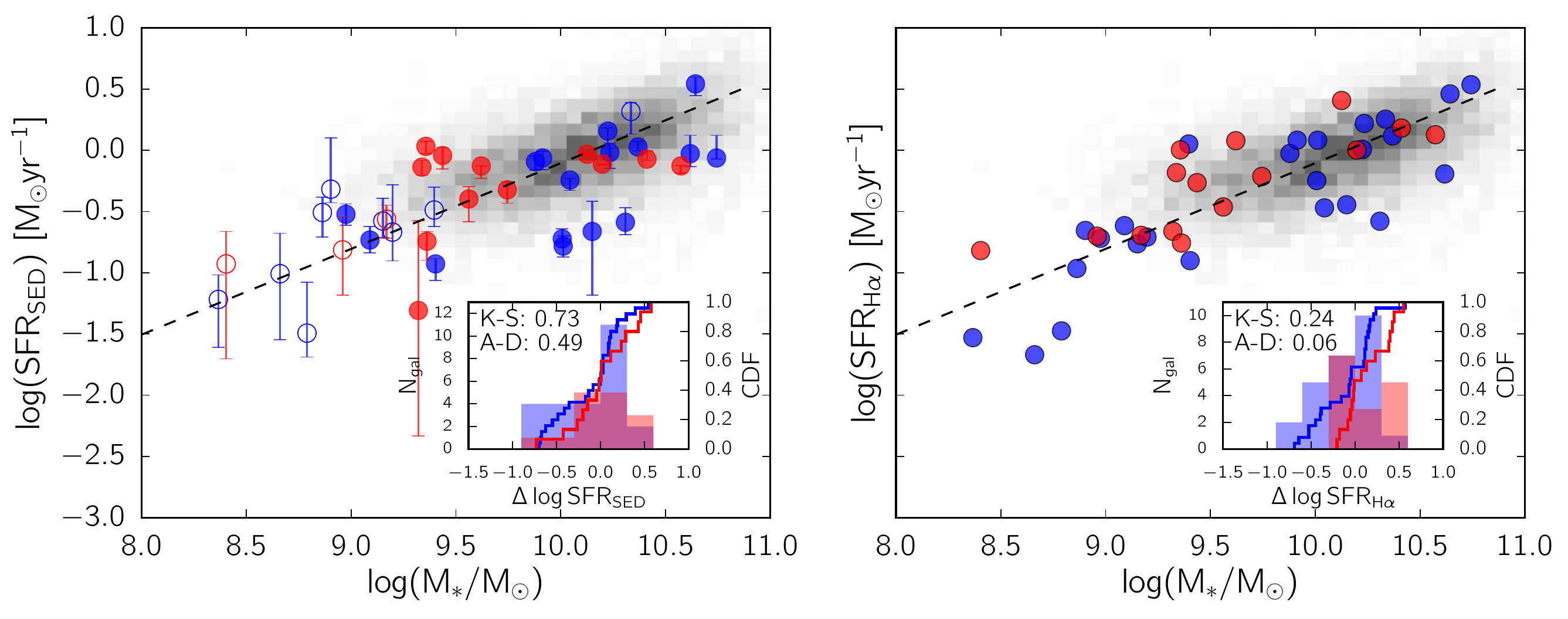}
\vspace{-0.5cm}
\caption{The wind-dominated galaxies (red) and the remaining galaxies (blue) in the star formation versus stellar mass plane. The left and right panels present the SFRs derived from SED fitting and H$\alpha$, respectively. The filled and open circles in the left panel correspond to galaxies with and without Herschel photometry.  The dashed lines indicate the galaxy main sequence defined by our H$\alpha$ SFRs. For comparison, the grey scales show distributions of star-forming edge-on discs (inclination angle $>80$ degrees or {\it b/a} $< 0.26$) in SDSS DR7 from the MPA/JHU catalogue. The insets show the distributions of $\Delta\log(\rm SFR)$, the difference between the measured SFR and that inferred from the dashed lines. The {\it p-}values of two-sample Kolmogorov-Smirnov test and Anderson-Darling test are labelled in the insets.  }\label{fig:corr_ms}
\end{figure*}

We now combine stellar masses and SFRs to investigate galactic winds on the galaxy main sequence, a log-linear relationship between stellar mass and SFR \citep{Noeske:2007lr}. Figure~\ref{fig:corr_ms} shows the two groups of galaxies on the galaxy main sequence plane. The dashed lines indicate the galaxy main sequence defined by our sample. We perform an unweighted linear fit to our data (using $\rm SFR_{H\mathnormal\alpha}$) while fixing the galaxy main sequence slope to well-defined slope of 0.7 \citep{,Zahid:2012ys,Noeske:2007lr}. For comparison, we also show edge-on star-forming discs selected from the SDSS MPA/JHU catalogues (inclination angle $>80$ degrees [{\it b/a} $< 0.26$] and $0.04<z<0.1$). We note that our main sequence is offset below that seen in the literature \citep[e.g.][]{Elbaz:2007dp,Zahid:2012ys} because the galaxy main sequence zero point depends on the observed axis ratio. The dashed-lines in Figure~\ref{fig:corr_ms} are only for visualization purposes, and our conclusions are not affected by the zero point of the galaxy main sequence.

Comparing with the SDSS galaxies, Figure~\ref{fig:corr_ms} demonstrates that our edge-on discs are indeed part of the normal star forming population in the local Universe. Although these main sequence galaxies have relatively modest SFRs, many of them are still capable of launching large-scale galactic winds that exhibit kinematic signatures resembling the bipolar winds observed in classical wind galaxies (Figure~\ref{fig:wind0}). Figure~\ref{fig:corr_ms} also suggests that the location of a galaxy on the galaxy main sequence plays little, if any, role in governing whether the galaxy hosts galactic winds. The insets of Figure~\ref{fig:corr_ms} show the distributions of $\Delta\log\rm SFR$, the difference between measured SFR and that inferred from the galaxy main sequence (dashed lines). The large {\it p-}values ($\gg0.2$) also suggest that the two sets of $\Delta\log\rm SFR_{SED}$ and $\Delta\log\rm SFR_{H\alpha}$ distributions are statistically similar. The only exception is the A-D {\it p-}value of 0.06 for the H$\alpha$ SFRs, which is only marginally higher than the threshold for significance. Interestingly, this may again indicate that galactic winds are more sensitive to star forming activities on short time-scales, consistent with the implication from sSFR in Figure~\ref{fig:corr_global}.

The lack of strong, if any, correlation with the galaxy main sequence is puzzling. The fact that normal galaxies follow the mass-size relation implies that galaxies lying above the galaxy main sequence for a given mass have higher SFRs and higher SFR surface densities than those below the galaxy main sequence \citep{Wuyts:2011fv}. While Figure~\ref{fig:corr_global} shows that SFR surface density correlates with the presence of winds, one would expect those high SFR surface density galaxies to occupy the upper part of the galaxy main sequence and therefore a correlation between the presence of winds and the location on the galaxy main sequence. The possibility of exhibiting such a correlation is suggested at a marginal level by the H$\alpha$ results. However, our data cannot place a firm conclusion due to systematic errors in SFRs, scatter in the mass-size relation and our small sample size.

\subsubsection{Star formation history}

\begin{figure}
\centering
\includegraphics[width = 8.5cm]{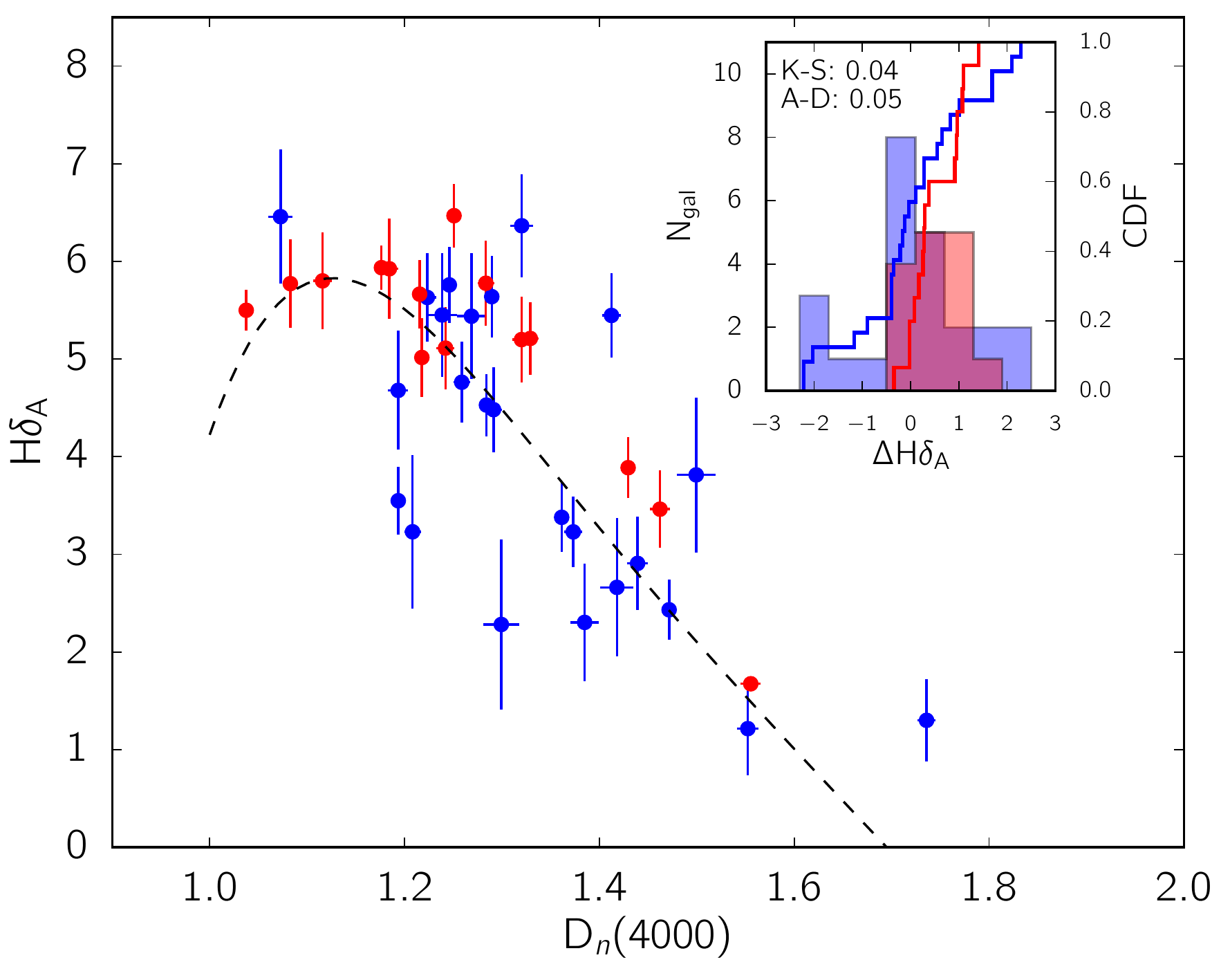}
\vspace{-0.5cm}
\caption{Star formation history traced by \Dn\ and \Hd\ of our galaxies. The wind-dominated galaxies are shown in red, and the remainder of the sample in blue. The dashed curve indicates the median SDSS relation. We fit a fifth order polynomial to the binned SDSS data in Figure~\ref{fig:sdss_sfh_sigma}. The inset shows the distributions of $\Delta$\Hd, the difference between observed \Hd and that inferred from \Dn\ given the median SDSS relation. The histograms correspond to the right y-axis, and the CDFs the left y-axis. The {\it p-}values of two-sample Kolmogorov-Smirnov test and Anderson-Darling test are labelled in the inset. }\label{fig:wind-sfh}
\end{figure}

Figure~\ref{fig:wind-sfh} compares the star formation history traced by \Dn\ and \Hd\ of the two groups of galaxies. The wind-dominated galaxies on average lie above the median \Dn-\Hd\ locus of SDSS (12 above and 3 below), whereas the remainder of the sample is distributed evenly above and below the SDSS median (12 above and 12 below). We further demonstrate this observation by showing in the inset of Figure~\ref{fig:wind-sfh} the histograms and cumulative distributions of $\Delta$\Hd, the difference between observed \Hd and that inferred from \Dn\ given the median SDSS relation. The mean $\Delta$\Hd\ for the wind-dominated galaxies and the remaining galaxies are 0.51\AA\ and 0.06\AA, respectively. The mean $\Delta$\Hd\ values can be compared with $0.23\pm0.21$\AA\ and $0.23\pm0.13$\AA\ (mean $\pm$ standard deviation), respectively, if the two populations were distributed randomly, implying that the difference in the mean $\Delta$\Hd\ is at about 2$\sigma$ significance. These mean and standard deviation values characterize the $\Delta$\Hd\ distributions resulting from randomly drawing the same numbers of galaxies (as the two groups) 10,000 times from our sample (i.e. Monte Carlo simulations). 

The Monte Carlo simulations also suggest that it is very unlikely to observe the configuration of the 15 wind-dominated galaxies, i.e.~12 above and 3 below the SDSS locus. The probability of having more than or equal to 12 galaxies above the SDSS locus is low (6\%) when randomly selecting 15 galaxies from our sample. Furthermore, the K-S and A-D tests both yield low probabilities ({\it p-}values of 0.04 and 0.05) for the two $\Delta$\Hd\ distributions to be drawn from the same parent distribution. We therefore conclude that the wind-dominated galaxies on average show enhanced \Hd\ indices, which presumably are driven by bursts of star formation occurring in the recent past. This conclusion is consistent with the implications in Sections 6.1.1. and 6.1.2 that the wind activities are more sensitive to star formation of shorter time-scales (i.e. H$\alpha$- versus SED-based SFRs in panels (c) and (g) in Figure~10, and Figure~11). We will uss the implications of these results in Section~7.1.

\subsection{Optical line ratios}
\begin{figure}
\centering
\includegraphics[width = 8.5cm]{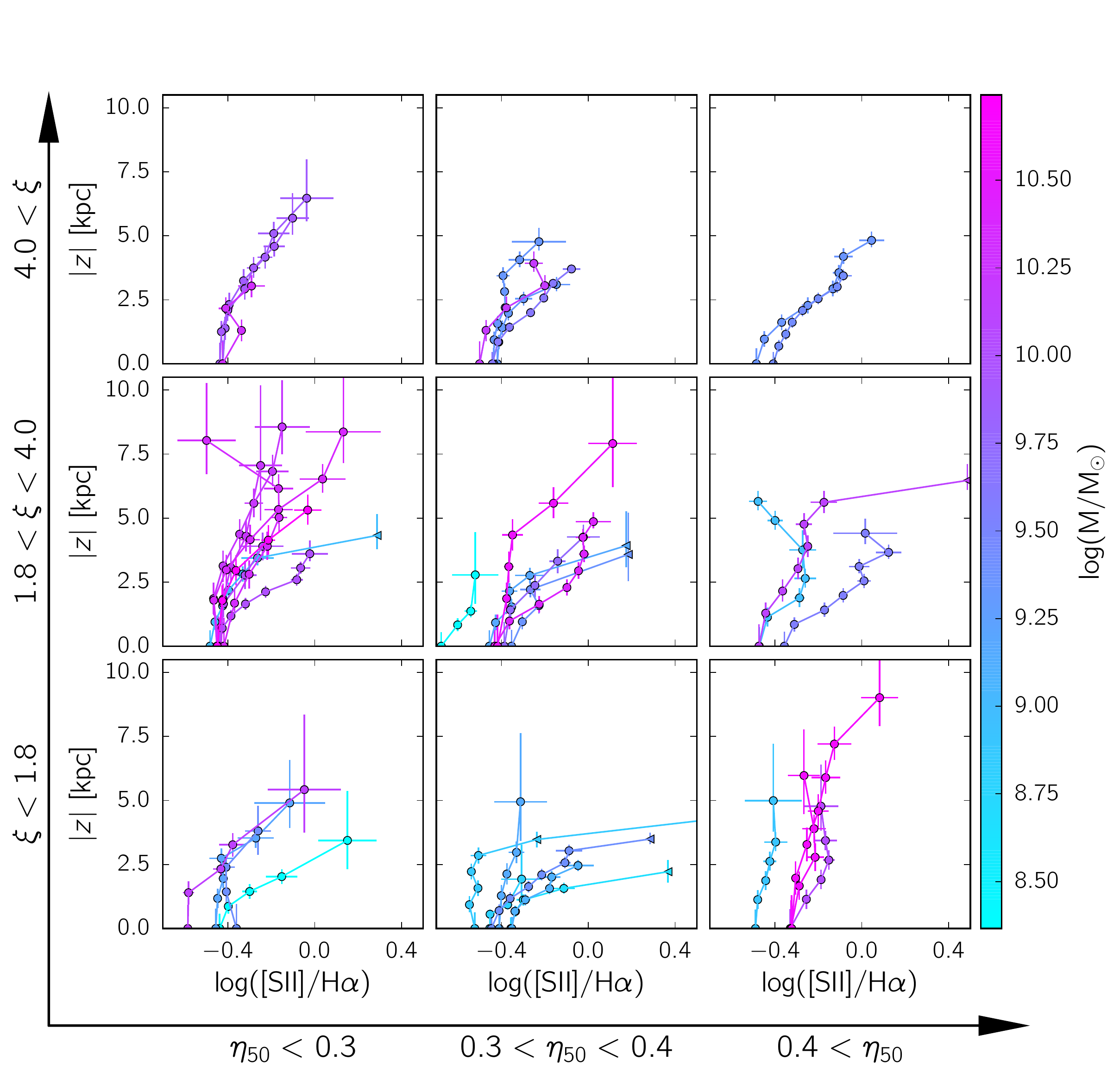}
\vspace{-0.3cm}
\caption{\SII/H$\alpha$ line ratio versus off-plane distance $|z|$. The 40 galaxies are presented in 9 different panels depending on their $\eta_{50}$ and $\xi$ values. The ranges of $\eta_{50}$ and $\xi$ each panel cover are labelled on the outer large axes (see also Figure~\ref{fig:xi-eta}). Each galaxy corresponds to one set of lines and the colour corresponds to the stellar mass as indicated in the colour bar. The line fluxes are measured with the adaptive $|z|$-binning approach described in Section~3.2. We apply a S/N cut of 3 on both lines and when the \SII\ line is not detected we calculate 3$\sigma$ upper limits assuming 100\kms\ line width. Triangles represent upper limits. The horizontal error bars are errors of the line ratio, and the vertical error bars indicate the $|z|$ range covered by each bin.}\label{fig:s2_profile}
\end{figure}

We now examine whether the kinematic properties of the extraplanar gas correlate with the extraplanar emission line ratios. In Figures~\ref{fig:31452} to \ref{fig:185510_residual}, we have shown that for galaxies with very different kinematics in their extraplanar gas the emission line ratios can have similar values and show similar increase with increasing $|z|$. Indeed, increase in the emission line ratios with increasing $|z|$ is common for both eDIG and galactic winds. Elevated \NII/H$\alpha$, \SII/H$\alpha$ and \OI/H$\alpha$ off the disc plane have been observed in the Milky Way and external galaxies that are not currently hosting large-scale galactic winds \citep[e.g.][]{Otte:2002fp,Miller:2003vn,Madsen:2006fj,Haffner:2009fr}. The enhanced line ratios are considered to be primarily caused by leaking photons from \HII\ regions photoionizing the low density, high temperature extraplanar gas, although additional heating sources such as turbulance, shocks, or cosmic rays could also be important \citep{Collins:2001bj,Miller:2003vn,Barnes:2014pi}. In classical wind systems \citep[e.g.][]{Sharp:2010qy}, these line ratios are also enhanced off the disc plane due to strong forbidden lines of low ionization species emitted by gas in the warm, partially ionized zone in the post-shock regions \citep{Dopita:1995fj,Dopita:1996uq,Allen:2008fk}. It is of interest to explore in our sample whether the increase in line ratios is different between galaxies with and without strong kinematic perturbations. 

\begin{figure}
\centering
\includegraphics[width = 8.5cm]{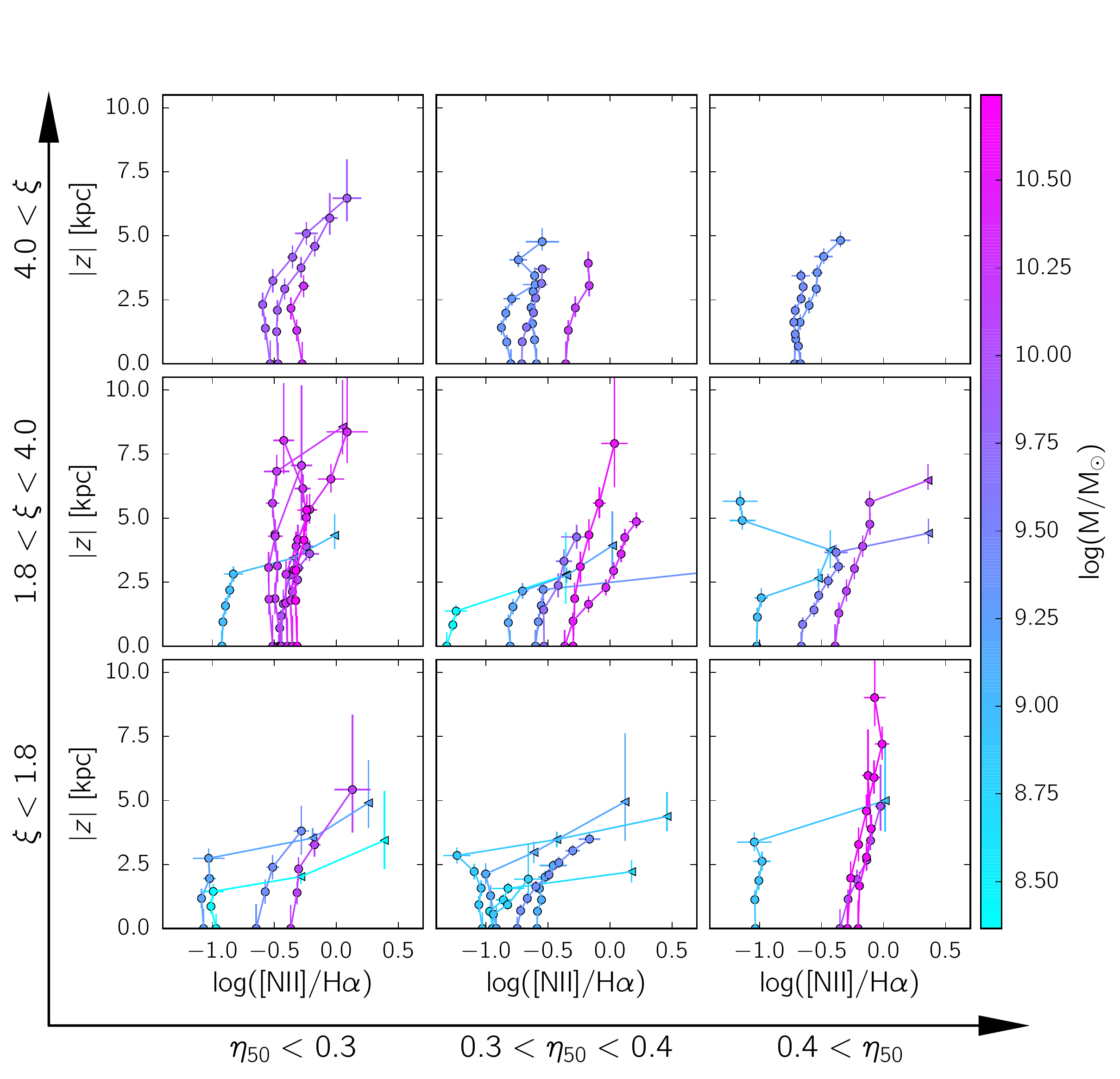}
\vspace{-0.3cm}
\caption{Same as Figure~\ref{fig:s2_profile} but for the \NII\ line.  }\label{fig:n2_profile}
\end{figure}

\begin{figure}
\centering
\includegraphics[width = 8.5cm]{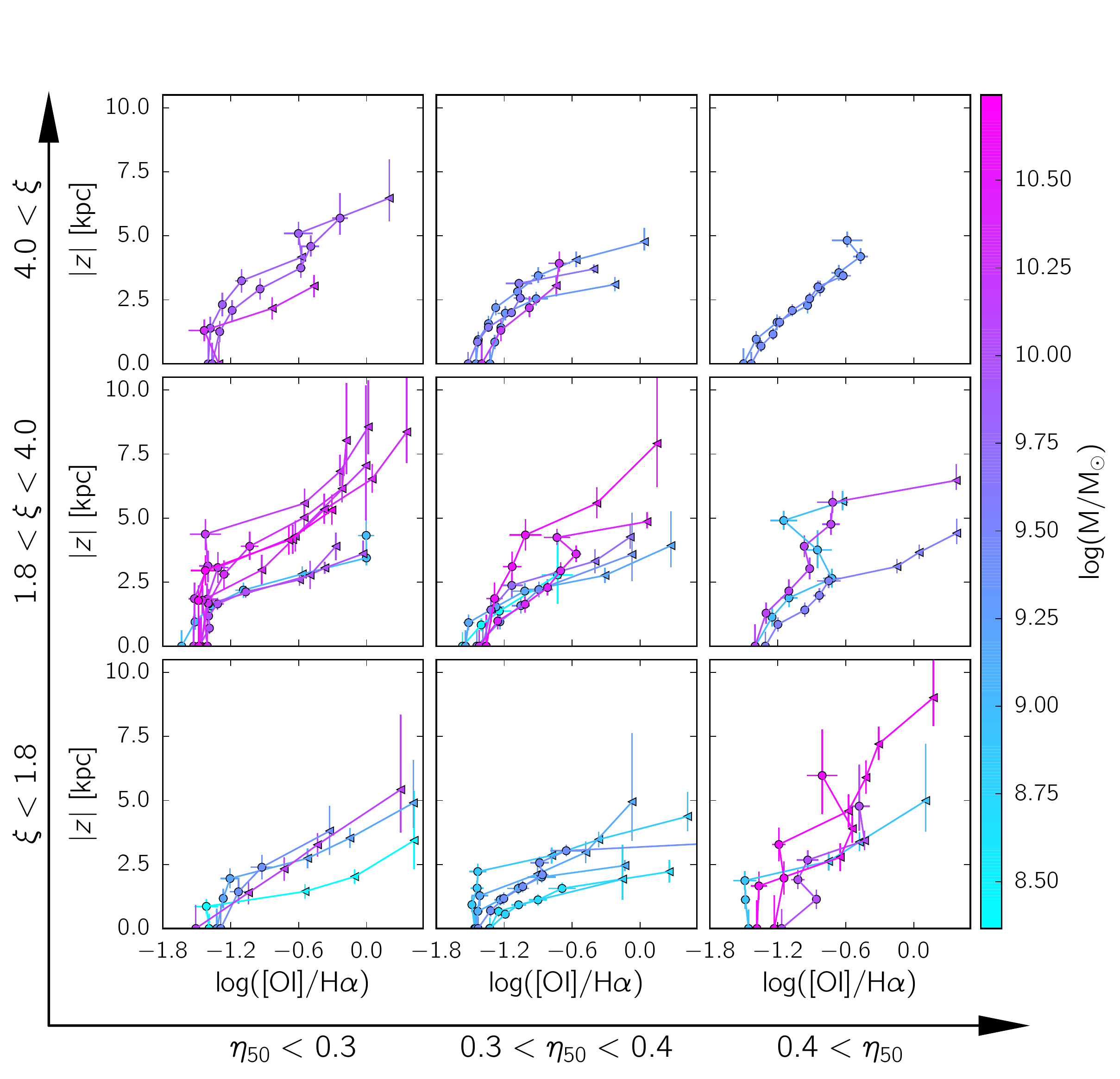}
\vspace{-0.3cm}
\caption{Same as Figure~\ref{fig:s2_profile} but for the \OI\ line.}\label{fig:o1_profile}
\end{figure}

We split our galaxies into 9 cells based on $\eta_{50}$ and $\xi$ and present in Figure~\ref{fig:s2_profile} their \SII/H$\alpha$ line ratio as a function of $|z|$. The line ratios are from the adaptive $|z|$-binning method. Here, we apply a S/N cut of 3 on both the \SII\ and H$\alpha$ lines, and when \SII\ is below 3$\sigma$ we estimate upper limits assuming a 100\kms\ line width. The line ratio \SII/H$\alpha$ has the advantage of having higher S/N than the weak \OI\ line and being less sensitive to the disc abundance and abundance gradient than the \NII/H$\alpha$ ratio. Figure~\ref{fig:s2_profile} shows that in virtually all galaxies the line ratio increases monotonically with $|z|$, and the increase shows no clear correlation with the kinematic properties of the extraplanar gas (i.e. $\eta_{50}$ and $\xi$). In other words, the wind-dominated galaxies show the same \SII/H$\alpha$ profiles in $|z|$ as those not dominated by galactic winds. Similar comparisons for the \NII/H$\alpha$ and \OI/H$\alpha$ line ratios are presented in Figures~\ref{fig:n2_profile} and \ref{fig:o1_profile}. These $|z|$ profiles also reveal no clear correlations between the line ratio profiles and the kinematic properties of the extraplanar gas. These results indicate that it can be misleading to use only the increase in strong optical line ratios with $|z|$ to infer the presence of either galactic winds or shocks. Even in galaxies with winds or shocks, eDIG could exist and may contaminate or dominate a correlation of line ratio with distance from the disc. Kinematic information such as either velocity asymmetry and/or elevated velocity dispersion must also be considered. 

\section{ussion}

\subsection{The roles of SFR surface density and star formation history in driving galactic winds}

The critical role that SFR surface density plays in driving galactic winds has long been recognized. The fact that galactic winds are detected ubiquitously in galaxies with $\rm \Sigma > 0.1\rm M_{\sun}~yr^{-1}~kpc^{-2}$ in early studies (see \citealt{Heckman:2002fk} for a review) has led to the common notion that there exists a ``threshold'' in SFR surface density above which winds are driven. The physics behind such a threshold is that this surface density could provide the energy input necessary to exceed the binding energy for the gas to escape into the halo \citep{Chevalier:1985fj,Dopita:2008qy}. The input energy could be thermal energy from supernova or radiation pressure from massive stars \citep{Murray:2005fk,Murray:2011fj}. The empirical threshold value is met by local starbursts, merging galaxies and high redshift galaxies, but not present-day spirals. The presence of galactic winds in more ``normal'' galaxies is now revealed by our integral field data and other recent studies \citep[e.g.][]{Chen:2010qy,Rubin:2014ly,Ho:2014uq}.

SFR surface density correlates with observed properties of starburst-driven winds, as revealed by several absorption studies using resonant lines (e.g. \ion{Na}{d}, \ion{Mg}{ii}, \ion{Fe}{ii}; \citealt{Rubin:2011yf}). \citet{Chen:2010qy} performed a stacking analysis on \ion{Na}{d} using SDSS spectra of present-day star-forming galaxies. They found that the equivalent width of the blueshifted (outflowing) component correlates strongly with SFR surface density. At higher redshifts, \citet{Bordoloi:2014vl} coadded the \ion{Mg}{ii} lines in $z \sim 1 \mbox{--}1.5$ galaxies and also found a strong correlation between the equivalent width and SFR surface density. \citet{Rubin:2010cs,Rubin:2014ly} found that such a correlation may still be present, although much less pronounced than others have indicated, in their studies of individual $z\sim0.5\mbox{--}1.5$ star-forming galaxies using \ion{Mg}{ii} and \ion{Fe}{ii} lines. The outflow velocity derived from the absorption line profiles also correlates with SFR surface density \citep{Kornei:2012zr,Bordoloi:2014vl}, especially when normalized by the galaxy circular velocity \citep{Heckman:2015if}. Our result of wind-dominated galaxies showing higher SFR surface density further emphasizes the importance of SFR surface density in driving galactic winds (Figure~\ref{fig:corr_global}).

While the importance of SFR surface density is well-established, SFR surface density is difficult to measure accurately, differs from definition to definition, and could depend on resolution. This makes the comparison of the various figures given in the literature rather nontrivial. The surface area over which star formation rate is divided exhibits a factor of two difference depending on whether one accounts for one side or both sides of the galaxy disc. In general, a characteristic scale radius is required to quantify the area over which the star formation activities spread. It is common to use the effective radius (adopted in this work), isophotal radius, or Petrosian radius. The characteristic scale radius should ideally be measured on a well-resolved rest-frame UV image to better quantify the size of the relevant star-forming disc. However, UV data can be subject to heavy extinction or could be simply unavailable; therefore, images taken at longer wavelengths are sometimes employed instead. In the latter case (including this work), the computed SFR surface density is a lower limit because the actively star-forming disc is typically smaller than the stellar disc that dominates at longer wavelengths. Furthermore, the clumpy-disc morphology at high redshifts raises a concern of over-estimating the star-forming area even when diffraction-limited UV images are available, because much of the area within the scale radius may have no star formation activity. This has motivated some authors to adopt a more sophisticated approach of defining a ``clump area'' that consists of only pixels above a certain SFR surface density threshold, i.e. counting only the star-forming area \citep{Kornei:2012zr}. While this idea is appealing, the defined area will certainly change with the threshold value and instrumental resolution. Evidently, the SFR surface density first proposed as a global, average quantity is scale-dependent, and depends critically on the area over which the measured star formation rate is averaged.

A second factor that could also control the presence of galactic winds is the star formation history. Our results suggest that large-scale winds trace previous bursts of star formation in the recent past (Section~6.1.3; Figure~\ref{fig:wind-sfh}). The potential connection between bursty star formation and galactic winds has already been suggested in a number of studies. \citet{Sato:2009ys} studied the \ion{Na}{d} lines at intermediate redshifts ($0.1<z<0.6$) with the DEEP2 survey and detected outflows in red-sequence galaxies. They found that many of these galaxies could have experienced recent starbursts as indicated by their enhanced H$\beta$ absorption lines (i.e. poststarburst signatures). \citet{Sharp:2010qy} studied 10 local wind galaxies with integral field spectroscopy and found that shock excitation dominates the extended wind-driven, line-emitting filaments in starburst galaxies. A simple time-scale argument demonstrates the need for bursty star formation to excite the filaments. \citet{Sharp:2010qy} compared the time evolution of ionizing and mechanical luminosities between an instantaneous burst model and a continuous star formation model. For shock excitation to dominate, the mechanical luminosity has to exceed the ionizing luminosity when the wind-driven filaments emerge from the disc. For continuous star formation, the ionizing luminosity dominates at all times, whereas for an instantaneous burst, the  mechanical luminosity exceeds the ionizing luminosity at about 10~Myr after the burst. This time-scale is comparable to the dynamical time for the filaments to develop (several Myr). \citet{Sharp:2010qy} concluded that there is a substantial delay (approximately 10~Myr) after the starburst before the conditions for a large-scale wind are properly established. 

The role of bursty star formation is also recognized in numerical simulations. \citet{Muratov:2015hl} analysed hydrodynamic cosmological zoom simulations from the Feedback in Realistic Environments (FIRE; \citealt{Hopkins:2014rz}) simulations and found that bursts of star formation are followed by powerful blasts of galactic outflows that could sweep gas out to the virial radius. They find that frequent bursts are the dominant form of star formation in all their simulated galaxies, particularly at high redshift. Present-day $L^*$ galaxies that have developed stable discs switch to a continuous mode of star formation and no longer drive large-scale outflows into the haloes, but the bursty form of star formation still persists in sub-$L^*$ galaxies.

\begin{figure}
\centering
\includegraphics[width = 8.5cm]{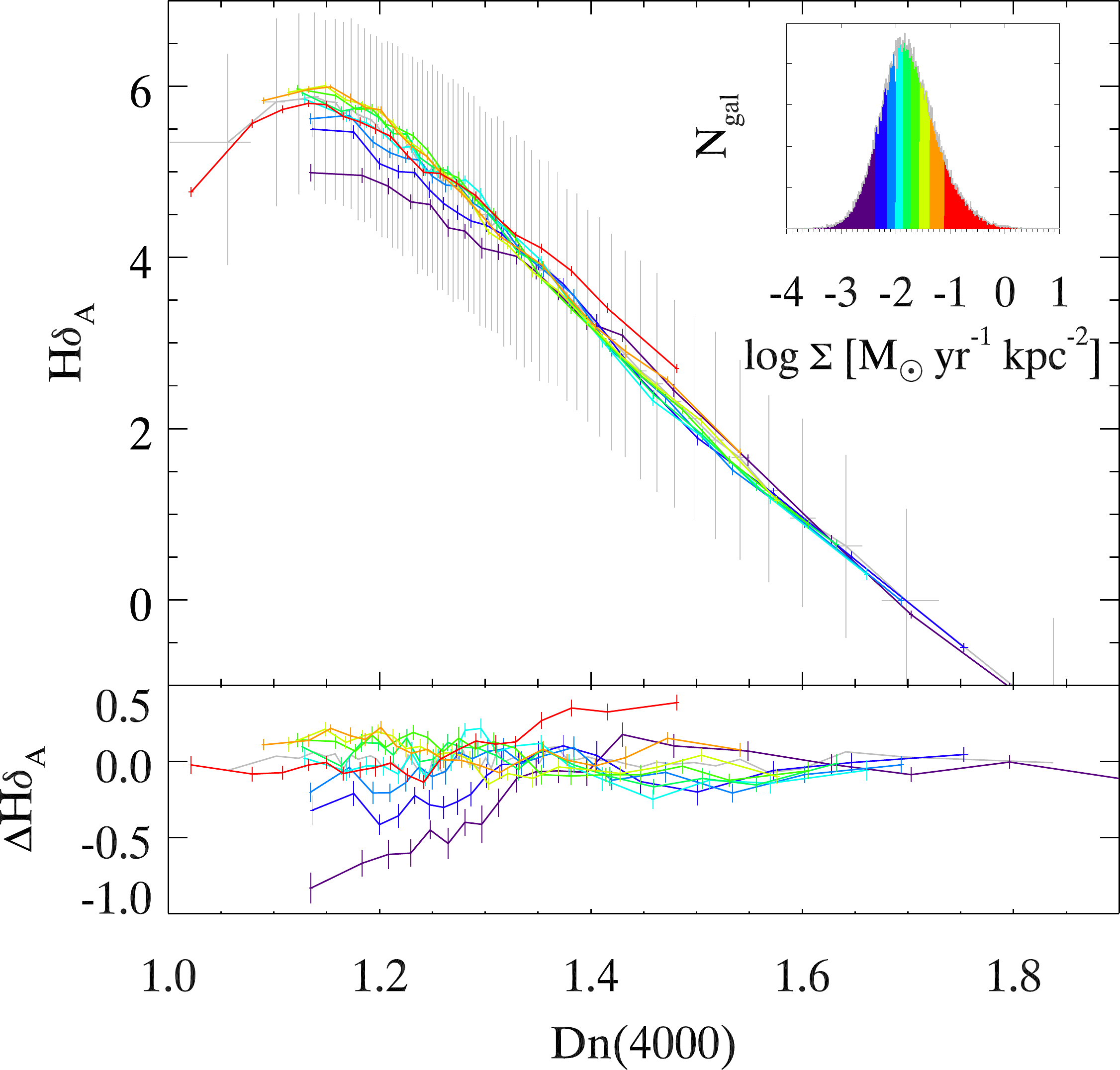}
\vspace{-0.5cm}
\caption{{\it Upper panel:} \Dn\ versus \Hd\ for SDSS DR7 star-forming galaxies ($0.04<z<0.1$ with non-AGN like \NII/H$\alpha$ and \OIII/H$\beta$ line ratios; \citealt{Kewley:2001lr}). The grey lines connect median values of all the star-forming galaxies ($\rm N_{gal}\approx112,000$) binned by \Dn. The (long) grey error bars contain 68\% of the data points in each bin. The colour lines connect median values of sub samples of galaxies binned by $\Sigma$ and \Dn. The corresponding (short) error bars are the bootstrapped $1\sigma$ errors of the median. Each colour bin contains approximately 620 galaxies. The inset indicates the distribution of $\Sigma$ and the different colours correspond to the different $\Sigma$ bins. {\it Lower panel:} Differences in \Hd\ between the whole sample and the $\Sigma$ bins ($\Delta$\Hd). A fifth order polynomial is fit to the whole sample (upper panel grey lines) to characterise the overall  \Hd(\Dn). All the values are taken from the MPA/JHU DR7 value-added catalogs (i.e. \Dn = {\tt D4000\_N\_SUB}; \Hd = {\tt LICK\_HD\_A\_SUB}). We adopt the same definition for SFR surface density as before, i.e. $\Sigma={\rm SFR}/2\pi r_e^2$ where $r_e$ is the {\it r}-band effective radius.}\label{fig:sdss_sfh_sigma}
\end{figure}

If the presence of galactic winds is related to both bursty star formation history and SFR surface density, then one expects a correlation between the two. This is indeed supported by existing SDSS data. Figure~\ref{fig:sdss_sfh_sigma} shows the median loci of SDSS star-forming galaxies on the \Dn\ versus \Hd\ plane binned by their SFR surface densities and \Dn. Figure~\ref{fig:sdss_sfh_sigma} clearly demonstrates that galaxies of different SFR surface densities have different star formation histories. Galaxies of higher SFR surface densities on average present higher \Hd\ values at a given \Dn\ than those with lower SFR surface densities. Excluding the first bin (red) with the highest SFR surface densities, the rest of the bins systematically decrease in \Hd, especially for $1.1\lesssim$\Dn$\lesssim1.3$. The small differences in \Hd\ are not surprising because 1) the delay time-scale of about 10~Myr corresponds to \Hd\ of approximately 5 to 7\AA, substantially smaller than the peak value of approximately 12 to 14\AA\ at a few hundred Myr to one Gyr, and 2) the starburst regions contribute only a fraction of the integrated light as measured by SDSS fibre. 
We will examine the latter with spatially resolved data in a future work. Despite the small differences in \Hd, the very small errors of the median (derived from bootstrapping) and the systematic trend all suggest that the SDSS data are entirely consistent with the picture that galaxies of high SFR surface densities have experienced bursts of star formation in the recent past. We expect these galaxies to host large-scale galactic winds. Which of the two physical characteristics is more important in governing the presence of large-scale winds remains to be tested in future works.

\subsection{Starburst- or AGN-driven winds?}

Determining whether the winds are driven by starburst or AGN activities is difficult, but we expect few AGNs in our sample. Only a small fraction ($\lesssim10\%$) of the emission line galaxies in SDSS contain optical AGNs \citep{Kewley:2001lr,Trouille:2010qv}. We also select our sample against optical AGNs using the nuclear optical line ratios (i.e.~[\ion{N}{ii}]/H$\alpha$ and [\ion{O}{iii}]/H$\beta$; \citealt{Kewley:2001lr}). Although the optical diagnostic could misidentify up to 50\% of X-ray AGNs as star-forming galaxies when classified using global spectra \citep{Trouille:2010qv,Pons:2014nr}, the small fraction of AGNs in emission line galaxies means that our sample is unlikely to contain more than four X-ray AGNs. Furthermore, we measure nuclear line ratios from the IFS data, and thus our classification is not subject to optical dilution caused by global emission line fluxes being contaminated by star formation activities in the host galaxies \citep{Moran:2002sf,Trump:2009mz}. Therefore, we believe that our winds are predominately starburst-driven. This hypothesis is consistent with the strong dependency between wind activities and SFR surface density which matches the theoretical predictions. However, it is impossible to completely rule out previous AGN activities driving the winds due to the rapid flickering of AGNs and the long fading time-scales of the filaments photoionized by the putative AGNs \citep{Novak:2011rt,Bland-Hawthorn:2013ty}.

\subsection{Accretion traced by the extraplanar gas}

While in this paper we predominately focus on connecting the extraplanar emissions to galactic winds and eDIG, the extraplanar gas could also be tracing gas accreted onto galaxies through satellite accretion. The best example is the Magellanic Stream seen in the Milky Way which extends more than 100 degrees away from the Magellanic Clouds and travels at approximately 100 to 200 \kms\ relatively to the Milky Way (in projection). Kinematic and morphological studies suggest that the Stream is directly related to tidal or hydrodynamical interactions between the Large and Small Magellanic Clouds and the Milky Way, during what could be  the clouds' first passage around the Milky Way \citep{Besla:2007fk}. The tip of the leading arm of the Stream has started to interact with the Milky Way disc as revealed by high-resolution \HI\ observations \citep{McClure-Griffiths:2008qf}. Along the Stream, H$\alpha$ emissions have also been detected and show fairly low velocity dispersions \citep[$\sigma<40$~\kms;][]{Weiner:1996rm}. The H$\alpha$ surface brightness ranging from a few tens to a few hundreds of milli-Rayleigh can be comparable to the other extraplanar gas off the Milky Way disc plane. Kinematically, the warm ionized gas is found to be associated with \HI\ gas, sharing the same velocity centroids \citep{Weiner:1996rm,Putman:2003rt,Madsen:2012fk}. It has been proposed that the \HI-dominated Stream is disrupted while travelling through the hot halo, resulting in shock cascade along the Stream that gives rise to the H$\alpha$ emission \citep{Bland-Hawthorn:2007lr,Bland-Hawthorn:2009fr,Tepper-Garcia:2015ul}. 

Satellite accretion events such as the Magellanic Clouds and Stream would naturally yield a high degree of velocity asymmetry (large $\xi$) but low off-plane velocity dispersion (small $\eta_{50}$).  In the case of the Milky Way, a high velocity asymmetry value is expected because there are no counterparts to the Magellanic Stream and Clouds north of the galactic plane. Although satellites can in principle accrete on both sides of the discs, it is unlikely that two accretion events (of similar mass and geometry) would occur at the same time on either side of the disc, and therefore we expect in general elevated $\xi$ when a satellite is accreted. The $\eta_{50}$ parameter, on the other hand, should not increase because the H$\alpha$ line in the Magellanic Stream shows fairly low velocity dispersions and such low velocity dispersions are expected in the shock cascade model \citep{Tepper-Garcia:2015ul}. Thus, the net effect of satellite accretion on our kinematic diagnostic diagram, Figure~\ref{fig:xi-eta}, is that galaxies with extraplanar gas dominated by satellite accretion would move to the upper left (i.e. high $\xi$ and low $\eta_{50}$). The existence of a population of such galaxies is suggested in Figure~\ref{fig:xi-eta}, under the caveat that the $\eta_{50}$ we measured are subject to the uncertainties of inferring $v_{rot}$ from the Tully-Fisher relation (see Sections 4.2 and 4.3). Follow-up high resolution, wide spatial coverage observations are necessary to study the nature of these kinds of systems.

\section{Summary and Conclusions}

We have examined extraplanar emissions in 40 edge-on disc galaxies using integral field spectroscopy data from the SAMI Galaxy Survey. These galaxies, spanning ranges in $\log(\rm M_*/M_{\sun})$ from 8.5 to 11 and $\log(\rm SFR/M_{\sun}~\rm yr^{-1})$ from  $-1.5$ to 1, are typical star-forming galaxies that fall on the galaxy main sequence. We have extracted gas velocity maps, velocity dispersion maps and emission line ratio maps from the imaging spectroscopy data to investigate physical properties of the extraplanar gas. Two parameters are invoked to quantify the kinematic properties of the extraplanar gas. The asymmetry parameter describes the asymmetry of the gas motion between the two sides of the disc. The velocity dispersion to rotation ratio parameter quantifies the level of increase in the velocity dispersion of the off-plane gas. These two parameters both increase if the extraplanar gas is predominately perturbed by strong disc-halo interactions through galactic winds, but the two parameters remain small if the disc interacts weakly with the halo, and the extraplanar gas co-rotates with the disc in the form of extended diffuse ionized gas. Using these two empirical parameters, we classify our sample into two groups: those with extraplanar gas dominated by galactic winds ($\rm N_{gal}=15$) and those not dominated by winds ($\rm N_{gal}=25$). By comparison of the physical properties of the two groups of galaxies we find that:
\begin{itemize}
\item{Galactic winds are preferentially detected in galaxies of high SFR surface densities. The wind-dominated galaxies present globally averaged SFR surface densities, i.e. SFR/$2\pi r_e^2$, below the canonical threshold of $0.1~\rm M_{\sun}~yr^{-1}~kpc^{-2}$.}
\item{Wind-dominated galaxies have enhanced \Hd\ compared to galaxies not dominated by galactic winds, implying that the two groups of galaxies have experienced different star formation histories. Enhanced \Hd\ is consistent with having experienced bursts of star formation in the recent past. }
\item{There are no strong trends with stellar mass, SFR, specific SFR, and locations of galaxies on the galaxy main sequence plane. There are indications of some trends with specific SFR and locations of galaxies on the galaxy main sequence such that galaxies with higher SFRs at a given mass are more likely to drive winds, but we are unable to draw firm conclusions using our data. }
\item{Emission line ratios (\NII/H$\alpha$, \SII/H$\alpha$, and \OI/H$\alpha$)} in general increase with increasing off-plane distance, showing no clear correlations with the kinematic properties of the extraplanar gas. 
\end{itemize}

Our results demonstrate that high SFR surface density and bursty star formation history are the two critical physical characteristics governing whether a galaxy is capable of developing large-scale galactic winds. The former is widely recognized in the literature and our results are consistent with a number of studies using other tracers to probe galactic winds. The latter is typically less constrained observationally although a simple time-scale argument concerning the excitation mechanism of the observed wind-driven filaments by \citet{Sharp:2010qy} has already implied that classical winds in nearby galaxies are driven by bursts of star formation. Recent hydrodynamic cosmological zoom simulations also suggest that frequent bursts of star formation dominating the star-forming activities are responsible for sweeping gas to the halo. This leads to the prediction that there should exist a correlation between SFR surface density and star formation history. We show that such correlation is indeed supported by existing SDSS data. SDSS galaxies of high SFR surface densities on average have enhanced \Hd, consistent with the picture that these galaxies have experienced bursts of star formation in the recent past and are likely to host large-scale winds. 

The major implications from this work are that 1) the so-called starburst-driven winds are indeed driven by bursts of star formation, and 2) probing star formation in the time domain is important for finding galactic winds and understanding how such disc-halo interactions affect the cycle of baryonic matter as galaxies evolve over cosmic time.

\section*{Acknowledgments}
We thank the anonymous referee for the thorough review and constructive comments. ITH thanks Ken Freeman, Elisabete da Cunha, Laura Zschaechner,  Catharine Wu, and George Heald for useful ussions. AMM, MAD and LJK acknowledge the support of the Australian Research Council (ARC) through overy project DP130103925. MAD would also like to thank the Deanship of Scientific Research (DSR), King Abdulaziz University for additional financial support as Distinguished Visiting Professor under the KAU Hi-Ci program. BG gratefully acknowledges the support of the ARC as the recipient of a Future Fellowship (FT140101202). JTA acknowledges the award of a SIEF John Stocker Fellowship. NB acknowledges the support of the EC FP7 SPACE project ASTRODEEP (Ref. No. 312725). JBH is funded by an ARC Laureate Fellowship. LC acknowledges financial support from the ARC (DP130100664). LD acknowledges support from the European Research Council Advanced Investigator grant, COSMICISM.

The SAMI Galaxy Survey is based on observations made at the Anglo-Australian Telescope. The Sydney-AAO Multi-object Integral field spectrograph was developed jointly by the University of Sydney and the Australian Astronomical Observatory. The SAMI input catalogue is based on data taken from the Sloan Digital Sky Survey, the GAMA Survey and the VST ATLAS Survey. The SAMI Galaxy Survey is funded by the Australian Research Council Centre of Excellence for All-sky Astrophysics, through project number CE110001020, and other participating institutions. The SAMI Galaxy Survey website is \href{http://sami-survey.org/}{http://sami-survey.org/}.

GAMA is a joint European-Australasian project based around a spectroscopic campaign using the Anglo-Australian Telescope. The GAMA input catalogue is based on data taken from the Sloan Digital Sky Survey and the UKIRT Infrared Deep Sky Survey. Complementary imaging of the GAMA regions is being obtained by a number of independent survey programmes including GALEX MIS, VST KiDS, VISTA VIKING, WISE, Herschel-ATLAS, GMRT and ASKAP providing UV to radio coverage. GAMA is funded by the STFC (UK), the ARC (Australia), the AAO, and the participating institutions. The GAMA website is \href{http://www.gama-survey.org/}{http://www.gama-survey.org/}.

The Herschel-ATLAS is a project with Herschel, which is an ESA space observatory with science instruments provided by European-led Principal Investigator consortia and with important participation from NASA. The H-ATLAS website is \href{http://www.h-atlas.org/}{http://www.h-atlas.org/}.

Funding for the Sloan Digital Sky Survey (SDSS) and SDSS-II has been provided by the Alfred P. Sloan Foundation, the Participating Institutions, the National Science Foundation, the U.S. Department of Energy, the National Aeronautics and Space Administration, the Japanese Monbukagakusho, and the Max Planck Society, and the Higher Education Funding Council for England. The SDSS Web site is \href{http://www.sdss.org/}{http://www.sdss.org/}.




\bibliographystyle{mnras}
\bibliography{/Users/itho/Dropbox/references}




\appendix
\section{Additional figures and table}
Figure~\ref{fig:nowind0} presents the 25 galaxies not classified as wind-dominated, i.e. $\eta_{50} < 0.3$ or $\xi < 1.8$ (see Section~4.3). The wind-dominated galaxies are presented in Figure~\ref{fig:wind0}. We summarise the basic properties of our sample in Table~\ref{table:sample}. 

\begin{figure*}
\includegraphics[width = 18cm]{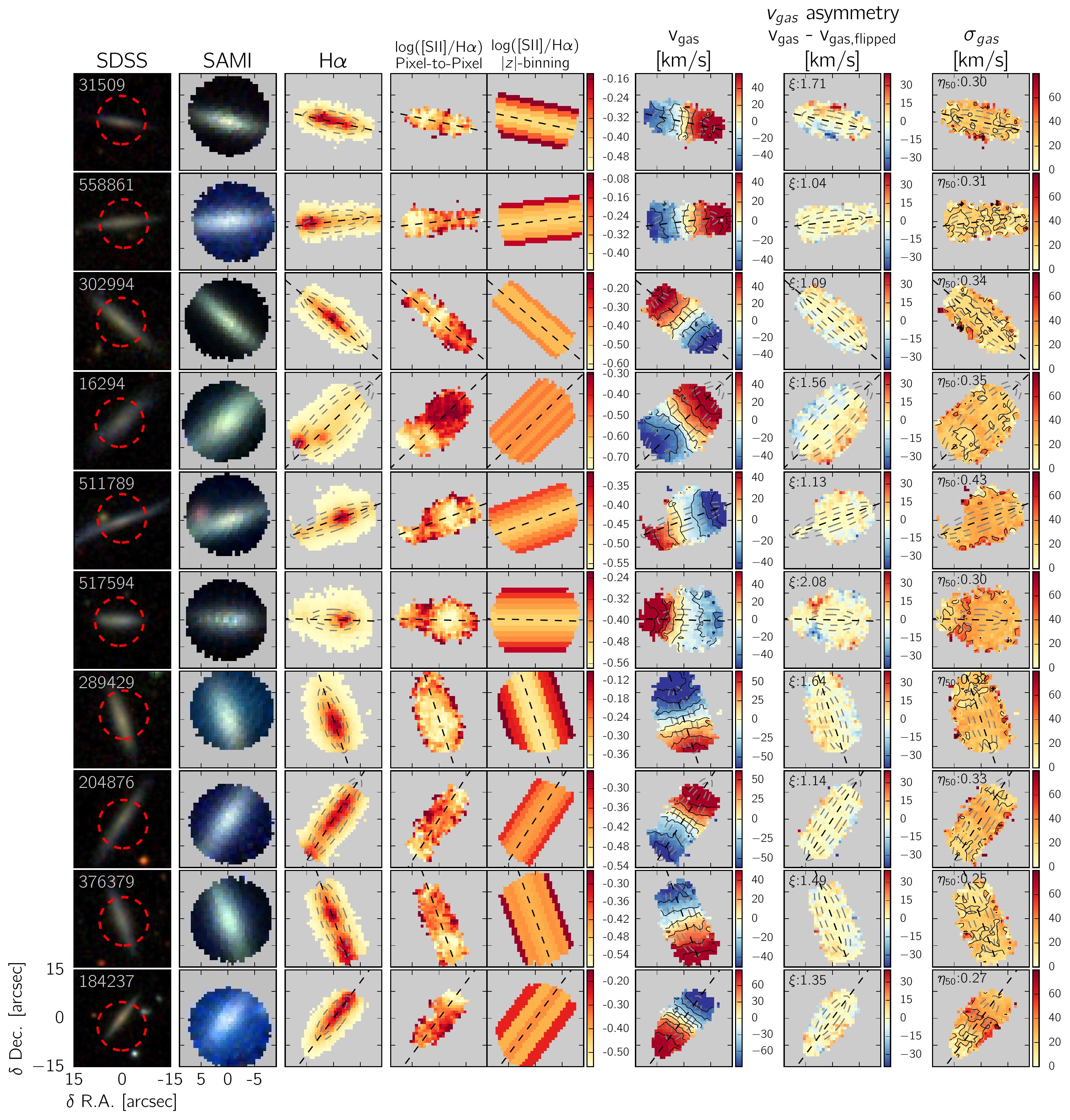}
\vspace{-0.5cm}
\caption{Same as Figure~\ref{fig:wind0} but for galaxies not dominated by galactic winds (i.e. $\eta_{50} < 0.3$ or $\xi < 1.8$).}\label{fig:nowind0}
\end{figure*}

\begin{figure*}
\includegraphics[width = 18cm]{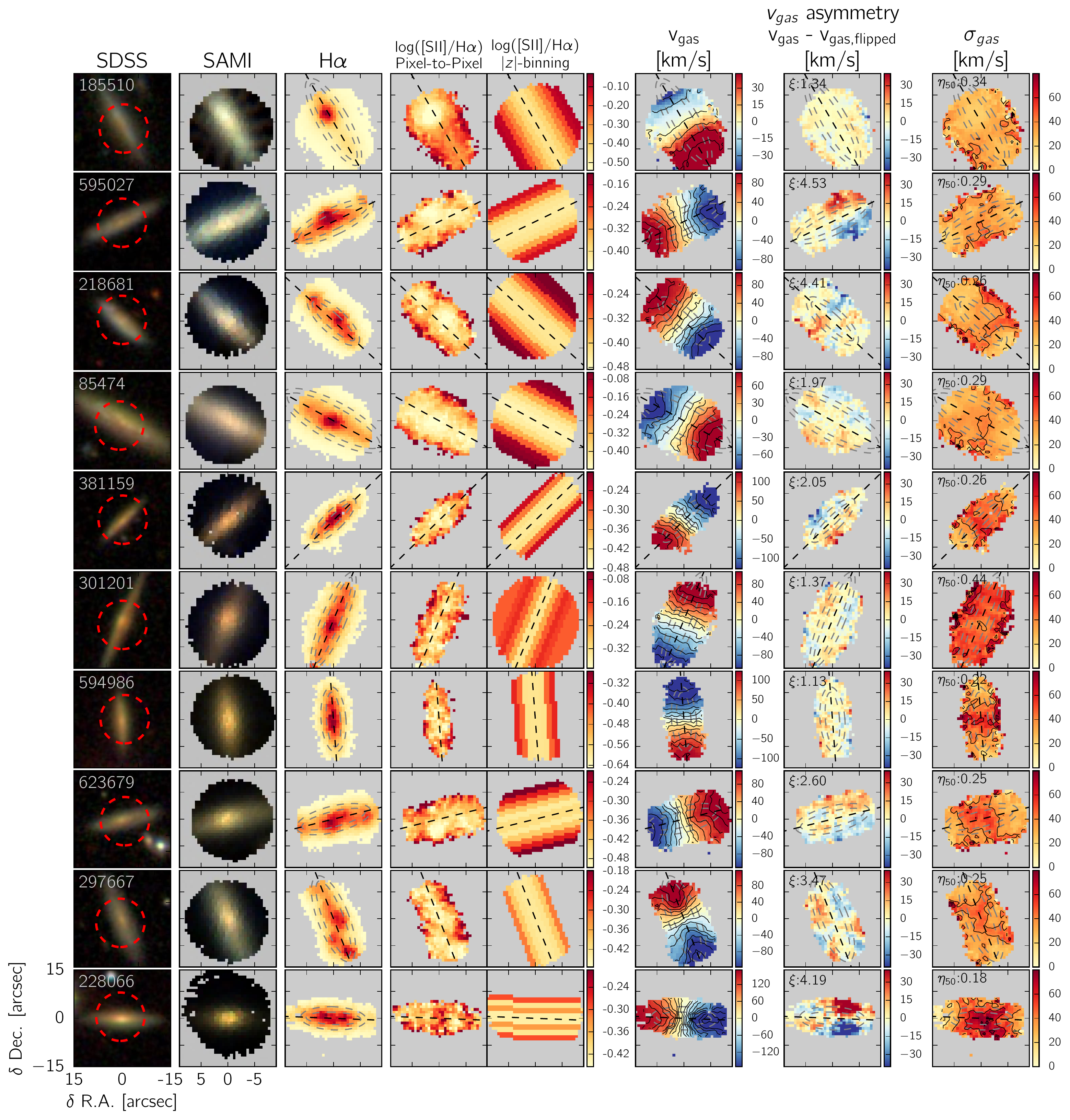}
\vspace{-0.5cm}
\contcaption{}\label{}
\end{figure*}

\begin{figure*}
\includegraphics[width = 18cm]{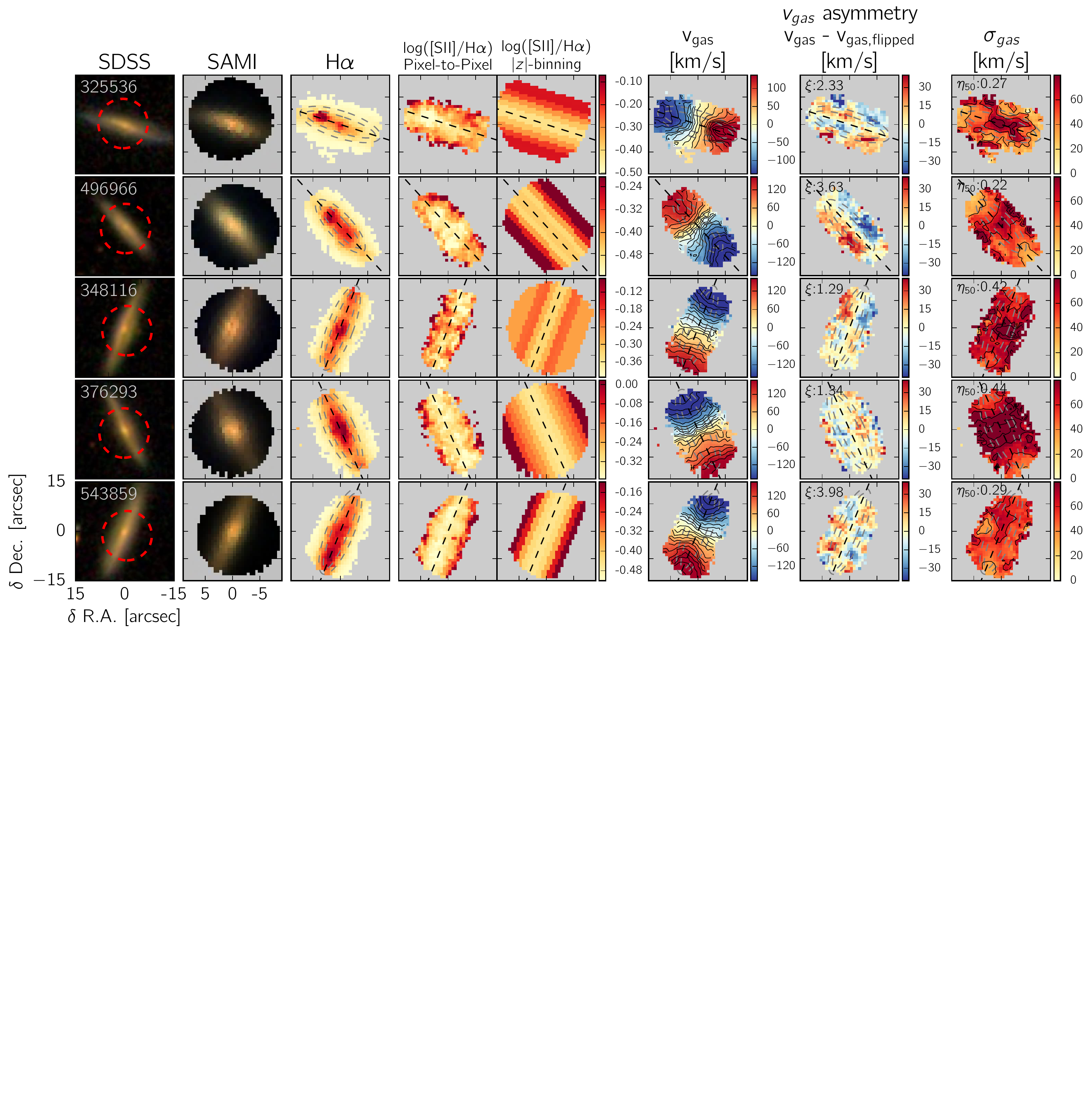}
\vspace{-8cm}
\contcaption{}\label{}
\end{figure*}


\begin{landscape}
 \begin{table}
  \caption{Sample properties}\label{table:sample}
  \label{tab:landscape}
  \begin{tabular}{lcccrrrccccccrl}
    \hline
    \multicolumn{1}{c}{GAMA} & R.A. & Dec. & $z^a$ & \multicolumn{1}{c}{$D^b$} & \multicolumn{1}{c}{$r_e$} & \multicolumn{1}{c}{$\log(M_*)$} & $\rm SFR_{SED}$ & $\rm SFR_{H\alpha}$ &$\rm\log\Sigma_{SED}$ & $\rm\log\Sigma_{H\alpha}$ & \Dn & \Hd & \multicolumn{1}{c}{$\xi$} & \multicolumn{1}{c}{$\eta_{50}^c$} \\
          \multicolumn{1}{c}{CATAID}        & [$^\circ$] &[$^\circ$] & -- & \multicolumn{1}{c}{[Mpc]} & \multicolumn{1}{c}{[\arcsec]/[kpc]} & \multicolumn{1}{c}{[$\rm M_\odot$]} & \multicolumn{2}{c}{[$\rm M_\odot~yr^{-1}$]} &  \multicolumn{2}{c}{[$\rm M_\odot~yr^{-1}~kpc^{-2}$]} & -- & [\AA] & \multicolumn{1}{c}{--} & \multicolumn{1}{c}{--} \\
    \hline
\multicolumn{15}{c}{wind-dominated}   \\
\hline
543769 & 212.8001 & $-0.9674$ & 0.0249 & 113.0 & 2.8/1.6 & 8.4 & $0.12^{+0.10}_{-0.10}$ & 0.15 & $-2.111$ & $-2.002$ & 1.12 & $5.8\pm 0.5$ & 2.57 & 0.31$^*$ \\
93167 & 219.2504 & $+0.5647$ & 0.0348 & 156.4 & 5.4/4.1 & 9.0 & $0.15^{+0.13}_{-0.09}$ & 0.20 & $-2.836$ & $-2.725$ & 1.18 & $5.9\pm 0.5$ & 2.36 & 0.44 \\
24433 & 185.7224 & $+1.2413$ & 0.0278 & 126.9 & 4.0/2.4 & 9.2 & $0.27^{+0.08}_{-0.08}$ & 0.20 & $-2.135$ & $-2.266$ & 1.28 & $5.8\pm 0.4$ & 2.15 & 0.31$^*$ \\
567624 & 212.5595 & $-0.5785$ & 0.0258 & 116.6 & 6.2/3.5 & 9.3 & $0.05^{+0.22}_{-0.04}$ & 0.22 & $-3.191$ & $-2.544$ & 1.25 & $6.5\pm 0.3$ & 4.11 & 0.32 \\
574200 & 134.5234 & $-0.0211$ & 0.0286 & 129.6 & 4.7/3.0 & 9.3 & $0.73^{+0.10}_{-0.10}$ & 0.66 & $-1.883$ & $-1.928$ & 1.22 & $5.0\pm 0.4$ & 5.31 & 0.40$^*$ \\
228432 & 217.3857 & $+1.1174$ & 0.0298 & 134.0 & 4.4/2.9 & 9.4 & $1.08^{+0.10}_{-0.14}$ & 1.01 & $-1.681$ & $-1.710$ & 1.08 & $5.8\pm 0.5$ & 4.09 & 0.43$^*$ \\
239249 & 217.0184 & $+1.6391$ & 0.0290 & 130.7 & 3.5/2.2 & 9.4 & $0.18^{+0.04}_{-0.05}$ & 0.17 & $-2.222$ & $-2.236$ & 1.24 & $5.1\pm 0.4$ & 1.96 & 0.33$^*$ \\
31452 & 179.8635 & $-1.1551$ & 0.0202 & 95.7 & 10.0/4.6 & 9.4 & $0.91^{+0.15}_{-0.20}$ & 0.54 & $-2.174$ & $-2.396$ & 1.04 & $5.5\pm 0.2$ & 11.88 & 0.49 \\
238125 & 213.3289 & $+1.6644$ & 0.0259 & 117.0 & 8.7/4.9 & 9.6 & $0.40^{+0.10}_{-0.14}$ & 0.34 & $-2.580$ & $-2.645$ & 1.22 & $5.7\pm 0.3$ & 1.88 & 0.41 \\
106616 & 216.7211 & $+0.9629$ & 0.0262 & 118.0 & 7.3/4.2 & 9.6 & $0.74^{+0.00}_{-0.15}$ & 1.20 & $-2.167$ & $-1.960$ & 1.18 & $5.9\pm 0.2$ & 4.58 & 0.35 \\
486834 & 221.7448 & $-1.7889$ & 0.0435 & 195.9 & 6.0/5.7 & 9.7 & $0.48^{+0.08}_{-0.11}$ & 0.61 & $-2.630$ & $-2.521$ & 1.33 & $5.2\pm 0.4$ & 2.51 & 0.34 \\
417678 & 132.7382 & $+2.3462$ & 0.0394 & 178.9 & 4.3/3.7 & 10.1 & $0.93^{+0.01}_{-0.01}$ & 2.55 & $-1.979$ & $-1.539$ & 1.32 & $5.2\pm 0.4$ & 3.48 & 0.56$^*$ \\
106389 & 215.9010 & $+1.0076$ & 0.0401 & 180.8 & 5.9/5.2 & 10.2 & $0.77^{+0.00}_{-0.01}$ & 1.00 & $-2.344$ & $-2.229$ & 1.46 & $3.5\pm 0.4$ & 5.03 & 0.32 \\
593680 & 217.4419 & $-0.1524$ & 0.0300 & 135.1 & 11.5/7.5 & 10.4 & $0.85^{+0.15}_{-0.02}$ & 1.52 & $-2.626$ & $-2.372$ & 1.43 & $3.9\pm 0.3$ & 2.67 & 0.40 \\
618906 & 217.3594 & $+0.3976$ & 0.0565 & 256.3 & 4.7/5.8 & 10.6 & $0.74^{+0.01}_{-0.10}$ & 1.34 & $-2.451$ & $-2.196$ & 1.56 & $1.7\pm 0.4$ & 2.45 & 0.31$^*$ \\
\hline
\multicolumn{15}{c}{not wind-dominated}   \\
\hline
31509 & 179.9708 & $-1.0899$ & 0.0260 & 119.4 & 4.8/2.8 & 8.4 & $0.06^{+0.04}_{-0.04}$ & 0.03 & $-2.907$ & $-3.219$ & 1.24 & $5.5\pm 0.6$ & 1.71 & 0.30$^*$ \\
558861 & 174.2570 & $-0.5120$ & 0.0194 & 92.2 & 5.6/2.5 & 8.7 & $0.10^{+0.11}_{-0.07}$ & 0.02 & $-2.602$ & $-3.264$ & 1.50 & $3.8\pm 0.8$ & 1.04 & 0.31 \\
302994 & 140.1432 & $+1.5511$ & 0.0167 & 76.4 & 7.1/2.6 & 8.8 & $0.03^{+0.05}_{-0.01}$ & 0.03 & $-3.134$ & $-3.115$ & 1.26 & $4.8\pm 0.4$ & 1.09 & 0.34 \\
16294 & 218.2074 & $+0.6370$ & 0.0290 & 130.4 & 9.0/5.7 & 8.9 & $0.31^{+0.10}_{-0.11}$ & 0.11 & $-2.812$ & $-3.270$ & 1.19 & $4.7\pm 0.6$ & 1.56 & 0.35 \\
511789 & 215.9925 & $-1.1284$ & 0.0344 & 154.9 & 7.4/5.6 & 8.9 & $0.48^{+0.78}_{-0.11}$ & 0.22 & $-2.609$ & $-2.946$ & 1.32 & $6.4\pm 0.5$ & 1.13 & 0.43 \\
517594 & 132.9978 & $+2.5434$ & 0.0285 & 129.3 & 3.9/2.4 & 9.0 & $0.30^{+0.06}_{-0.06}$ & 0.19 & $-2.095$ & $-2.291$ & 1.07 & $6.5\pm 0.7$ & 2.08 & 0.30$^*$ \\
289429 & 182.3047 & $+1.8366$ & 0.0194 & 92.5 & 5.5/2.5 & 9.1 & $0.18^{+0.05}_{-0.04}$ & 0.24 & $-2.314$ & $-2.197$ & 1.21 & $3.2\pm 0.8$ & 1.64 & 0.32 \\
204876 & 139.8844 & $-0.4101$ & 0.0387 & 175.7 & 8.0/6.8 & 9.2 & $0.26^{+0.14}_{-0.07}$ & 0.17 & $-3.046$ & $-3.232$ & 1.27 & $5.4\pm 0.6$ & 1.14 & 0.33 \\
376379 & 133.0553 & $+1.4996$ & 0.0357 & 161.8 & 6.4/5.0 & 9.2 & $0.21^{+0.31}_{-0.09}$ & 0.20 & $-2.863$ & $-2.904$ & 1.22 & $5.6\pm 0.5$ & 1.49 & 0.25 \\
184237 & 175.7381 & $-1.5391$ & 0.0434 & 197.7 & 4.0/3.8 & 9.4 & $0.33^{+0.17}_{-0.09}$ & 1.13 & $-2.453$ & $-1.914$ & -- & -- & 1.35 & 0.27$^*$ \\
185510 & 180.5913 & $-1.4539$ & 0.0205 & 96.6 & 8.2/3.8 & 9.4 & $0.12^{+0.01}_{-0.03}$ & 0.13 & $-2.895$ & $-2.868$ & 1.25 & $5.8\pm 0.4$ & 1.34 & 0.34 \\
595027 & 223.1331 & $-0.1664$ & 0.0425 & 191.1 & 7.2/6.6 & 9.9 & $0.81^{+0.09}_{-0.12}$ & 0.94 & $-2.534$ & $-2.466$ & 1.29 & $5.6\pm 0.4$ & 4.53 & 0.29 \\
218681 & 139.2505 & $+0.8441$ & 0.0380 & 172.5 & 5.2/4.4 & 9.9 & $0.86^{+0.12}_{-0.17}$ & 1.21 & $-2.140$ & $-1.995$ & 1.28 & $4.5\pm 0.3$ & 4.41 & 0.26$^*$ \\
85474 & 182.5694 & $+0.6266$ & 0.0207 & 97.4 & 9.7/4.6 & 10.0 & $0.19^{+0.04}_{-0.01}$ & 0.57 & $-2.845$ & $-2.369$ & 1.19 & $3.5\pm 0.3$ & 1.97 & 0.29 \\
\hline
  \end{tabular}
    \begin{tablenotes}
      \small
      \item $^a$Redshift
      \item $^b$Luminosity distance inferred from redshift after correcting for the local and large-scale flows \citep{Tonry:2000lr}.
      \item $^c$Those with asterisk ($^*$) have $v_{rot}$ measured from the data. The others have $v_{rot}$ inferred from the Tully-Fisher relation (see Section~4.2). 
    \end{tablenotes}
 \end{table}
\end{landscape}

\begin{landscape}
 \begin{table}
  \contcaption{Sample properties}
  \label{tab:landscape}
  \begin{tabular}{lcccrrrccccccrl}
    \hline
    \multicolumn{1}{c}{GAMA} & R.A. & Dec. & $z^a$ & \multicolumn{1}{c}{$D^b$} & \multicolumn{1}{c}{$r_e$} & \multicolumn{1}{c}{$\log(M_*)$} & $\rm SFR_{SED}$ & $\rm SFR_{H\alpha}$ &$\rm\log\Sigma_{SED}$ & $\rm\log\Sigma_{H\alpha}$ & \Dn & \Hd & \multicolumn{1}{c}{$\xi$} & \multicolumn{1}{c}{$\eta_{50}^c$} \\
          \multicolumn{1}{c}{CATAID}        & [$^\circ$] &[$^\circ$] & -- & \multicolumn{1}{c}{[Mpc]} & \multicolumn{1}{c}{[\arcsec]/[kpc]} & \multicolumn{1}{c}{[$\rm M_\odot$]} & \multicolumn{2}{c}{[$\rm M_\odot~yr^{-1}$]} &  \multicolumn{2}{c}{[$\rm M_\odot~yr^{-1}~kpc^{-2}$]} & -- & [\AA] & \multicolumn{1}{c}{--} & \multicolumn{1}{c}{--} \\
    \hline
381159 & 131.6598 & $+1.7285$ & 0.0505 & 229.4 & 4.6/5.1 & 10.0 & $0.16^{+0.04}_{-0.03}$ & 1.21 & $-2.990$ & $-2.125$ & 1.30 & $2.3\pm 0.9$ & 2.05 & 0.26$^*$ \\
301201 & 132.7204 & $+1.2501$ & 0.0348 & 157.5 & 8.6/6.5 & 10.0 & $0.57^{+0.02}_{-0.10}$ & 0.34 & $-2.673$ & $-2.900$ & 1.42 & $2.7\pm 0.7$ & 1.37 & 0.44 \\
594986 & 222.8160 & $-0.1633$ & 0.0428 & 192.5 & 5.1/4.7 & 10.2 & $0.22^{+0.16}_{-0.15}$ & 0.36 & $-2.813$ & $-2.595$ & 1.38 & $2.3\pm 0.6$ & 1.13 & 0.22$^*$ \\
623679 & 139.9831 & $+0.6413$ & 0.0564 & 257.2 & 6.1/7.6 & 10.2 & $1.44^{+0.09}_{-0.40}$ & 1.02 & $-2.399$ & $-2.548$ & 1.41 & $5.4\pm 0.4$ & 2.60 & 0.25 \\
297667 & 216.7625 & $+1.3627$ & 0.0558 & 253.3 & 7.5/9.2 & 10.2 & $0.96^{+0.18}_{-0.25}$ & 1.66 & $-2.744$ & $-2.507$ & 1.29 & $4.5\pm 0.4$ & 3.47 & 0.25 \\
228066 & 215.8674 & $+1.1021$ & 0.0397 & 178.9 & 5.1/4.4 & 10.3 & $0.26^{+0.08}_{-0.05}$ & 0.26 & $-2.678$ & $-2.669$ & 1.74 & $1.3\pm 0.4$ & 4.19 & 0.18$^*$ \\
325536 & 140.9840 & $+1.8344$ & 0.0507 & 230.9 & 6.7/7.4 & 10.3 & $2.07^{+0.36}_{-0.72}$ & 1.79 & $-2.224$ & $-2.287$ & 1.36 & $3.4\pm 0.4$ & 2.33 & 0.27 \\
496966 & 212.5919 & $-1.1150$ & 0.0542 & 245.8 & 5.3/6.3 & 10.4 & $1.06^{+0.12}_{-0.07}$ & 1.31 & $-2.366$ & $-2.276$ & 1.44 & $2.9\pm 0.5$ & 3.63 & 0.22$^*$ \\
348116 & 140.2934 & $+2.2012$ & 0.0504 & 229.4 & 7.2/8.0 & 10.6 & $0.94^{+0.39}_{-0.20}$ & 0.64 & $-2.632$ & $-2.797$ & 1.55 & $1.2\pm 0.5$ & 1.29 & 0.42 \\
376293 & 132.7976 & $+1.5017$ & 0.0595 & 271.0 & 7.0/9.1 & 10.6 & $3.48^{+0.47}_{-0.68}$ & 2.88 & $-2.175$ & $-2.257$ & 1.37 & $3.2\pm 0.4$ & 1.34 & 0.44 \\
543859 & 213.1517 & $-1.0364$ & 0.0538 & 244.3 & 7.4/8.8 & 10.7 & $0.86^{+0.46}_{-0.02}$ & 3.43 & $-2.747$ & $-2.148$ & 1.47 & $2.4\pm 0.3$ & 3.98 & 0.29 \\
\hline

  \end{tabular}
 \end{table}
\end{landscape}





\bsp	
\label{lastpage}
\end{CJK*}
\end{document}
